\documentclass{PoS}
\usepackage{subfigure}
\usepackage{cite}
\title{Cosmic rays: direct measurements}
\ShortTitle{Cosmic rays: direct measurements}

\author{Paolo Maestro\\ 
        Department of Physical Sciences, Earth and Environment\\
        University of Siena, via Roma 56, 53100 Siena (Italy)\\
        E-mail: {paolo.maestro@pi.infn.it} 
}
\abstract{
This paper is based  on the rapporteur talk 
given at the 34$^{th}$ International Cosmic Ray Conference (ICRC),  on August 6$^{th}$, 2015. 
The purpose of the talk and paper is to provide a summary of the  most recent results 
from balloon-borne and space-based experiments
presented at the conference, 
and give an overview of the future missions and developments foreseen in this field. 
}

\FullConference{The 34th International Cosmic Ray Conference,\\
		30 July- 6 August, 2015\\
		The Hague, The Netherlands}
\begin{document}
\section{Introduction}
I will report on the papers concerning the direct measurements of cosmic rays
 (CRs) presented at the  34$^{th}$ ICRC, held in The Hague from July 30$^{th}$ to August 6$^{th}$, 2015.
I identified about 110 contributions relevant for this topic,  
mainly presented in the CR sessions (CR-EXperiments (45),
CR-INstrumentation (23), CR-THeory (25)) and Dark Matter sessions (DM-IN (10), DM-EX (2), DM-TH (4)).
Since it would be impossible to review all of them,  I 
 tried to highlight the ones I consider of interest for the broad CR community 
and which in my opinion better represent the status of the art in the field. 
In spite of my effort to be as much unbiased as possible,  
I realize that my judgement in this matter could be imperfect and the choice of topics inevitably affected by 
my research interests and experimentalist's point of view, and therefore I apologize for 
those contributions I omitted to mention. 

A wealth of new results obtained by balloon-borne 
and space-based experiments have been presented at the conference, covering 
several observation targets from nuclei and isotopes, to electrons and positrons, to antiprotons, measured 
over six decades in energy from few MeV up to some TeV.
The measurements from the AMS-02 and PAMELA spectrometers  are characterized by a precision never reached before
in CR history; they  allowed to discover features in the CR energy spectra 
that compel a revision of the simple idea that CR fluxes are described by smooth single power-laws in energy, 
and prompted an intense theoretical activity to interpret the results. 
Among the most interesting outcomes, 
the first measurement of the CR spectra outside the heliosphere (Voyager 1), 
the high-statistics measurement of the abundances of nuclei heavier than iron (SuperTIGER), 
and the first detection of a primary CR clock (ACE-CRIS)
are certainly  worth considering. 
Moreover upcoming calorimetric missions (NUCLEON, CALET, DAMPE, ISS-CREAM) 
are foreseen to collect in the next few years unprecedented statistics of CRs
even at higher energies, in the multi-TeV range and finally approaching the knee.
All these advances indicate that we have entered 
a new era of precision measurements of CRs, 
which can contribute to shed light on several fundamental questions in CR physics, 
still open a century after their discovery, 
like the mechanism of acceleration of galactic cosmic rays (GCRs), the nature and composition of their
sources, the CR propagation in the interstellar medium (ISM).

This paper is organized as the rapporteur talk, 
i.e. according to observation targets.  
In section \ref{s:pHe}, the most recent measurements of proton and helium are discussed, followed by 
results on heavier nuclei  in section \ref{s:nuclei}.
Section \ref{s:electron} deals with electrons, while 
measurements of the antimatter component (e$^+$, $\bar{p}$) of CRs are covered in section \ref{s:antimatter}. 
In each section, I briefly summarize  the main theoretical models which have been
proposed to interpret the results, though a complete discussion of this subject
 is beyond the scope of the paper. 
A review about the upcoming missions and proposals of new instrumentation and experiments is presented in section \ref{s:futurexp}.  
My conclusions are drawn in section \ref{s:conclusion}.
\section{Proton and helium}
\label{s:pHe}
Direct measurements of the energy and charge
of CR nuclei are performed by space-based or balloon-borne instruments exploiting various detection
techniques, depending on the energy region of interest. 
Stack of solid-state detectors are used up to hundreds of MeV/n, Cerenkov detectors up to few GeV/n, 
magnetic spectrometers in the energy range from few GeV/n up to few TeV/n, nuclear emulsions, calorimeters, transition
radiation detectors up to hundreds of TeV. 
Protons and helium nuclei are the most abundant components of cosmic radiation. 
Precise measurements of their fluxes at high energies can provide new insights into the nature and origin of CRs, 
while low-energy data are important for studying the CR 
interactions with ISM, heliosphere and magnetosphere.
\subsection{High-energy region}
The highest-energy data for proton and helium were obtained in the 1990's by balloon
experiments using nuclear emulsions (RUNJOB, JACEE), but they are affected by large errors and
inconsistent with each other \cite{RUNJOB,JACEE}. Measurements from JACEE suggested a steeper spectrum for $p$
with respect to He, but subsequent data obtained by RUNJOB did not confirm this result and were consistent
with a similar spectral index for the two lightest elements. This controversy was solved by the balloon-borne
experiment CREAM that collected the largest statistics of CR protons and helium nuclei above 100 TeV, 
showing unequivocally 
that $p$ and
He spectra in the multi-TeV energy region are harder than the energy spectra at lower energies 
($<$100 GeV/n) and described by power-laws $E^{\gamma}$ with different
spectral indices
($\gamma_p$= -2.66$\pm$0.02 and $\gamma_{He}$= -2.58 $\pm$0.02 in the particle energy range 2.5-250 TeV), with the proton softer than the He spectrum \cite{CREAMpHe}.
Then the only way to reconcile the higher and lower energy data is assuming a hardening of the
spectra in the hundred GeV region. 
This was first observed  by the magnetic spectrometer PAMELA that reported spectral shapes for $p$ and He clearly inconsistent
with single power-laws. PAMELA $p$ and He data
exhibit an abrupt spectral hardening occurring between 230 and 240 GV,
with a change $\Delta\gamma$ of the spectral index $\sim$0.2 for $p$ and 0.3 for He, respectively \cite{PAMELApHe}.\\
At this ICRC, AMS presented its final results on $p$ and He  confirming the break 
and providing a precise measurement of the transition to harder spectra.  
AMS-02 is a precision, multipurpose spectrometer launched in 2011 and installed on the International Space Station (ISS) \cite{AMS02}. 
The core of the instrument is a 0.14 T permanent magnet instrumented with nine layers of Si microstrip detectors, six of which are within the bore of the magnet.
The tracker accurately determines the trajectory
of charged particles in the magnetic field and measures the particle rigidity  $R=P/Z$ ($P$ is the momentum
and $Z$ the charge of the particle) from its curvature. 
The spatial resolution in each layer is 10 (7.5) $\mu$m for $p$ (He) in the bending direction, and  
the maximum detectable rigidity (MDR), where the error of the measured rigidity 
becomes 100\%, is 2 TV for $p$ and 3.2 TV for He, respectively. 
The tracker can also determine the absolute value of $Z$ by multiple dE/dx with 
a charge resolution $\Delta Z\approx$0.05 for |$Z$|=1 and 0.07 for |$Z$|=2.
In addition to the tracker, 
independent and redundant charge and energy measurements are provided by 
Time of Flight counters (TOF), a Ring Image Cerenkov Counter (RICH), and an Electromagnetic Calorimeter (ECAL).
A Transition Radiation Detector (TRD) is used to separate electrons and from protons,
while a veto system (ACC) rejects with an
efficiency of 0.99999 CRs which enter the inner tracker
from the side.\\
In four years of data taking AMS-02 collected about 68$\times$10$^9$  CR events, more than the whole statistics of all the previous 
balloon-borne and space-based experiments. 
A final selection of 300$\times$10$^6$ protons and 50$\times$10$^6$ He nuclei
was used to measure the differential fluxes as a function of rigidity, in the range  
between 1 GV and 1.8 TV for $p$ \cite{AMS-proton, AMS-protonPRL}, and from 1.9 GV to 3 TV for He \cite{AMS-helium}.
Such a huge statistics allowed a very accurate study of the systematic uncertainties affecting the flux measurements, 
comparable to the sophisticated analyses of collider experiments. 
The total systematic error includes  uncertainties 
stemming from several sources, like 
the trigger efficiency, the acceptance, the background contamination, the geomagnetic cutoff factor,
the event selection, the unfolding, the rigidity resolution function, and the absolute rigidity scale.
Systematic errors dominate
in whole rigidity range, the statistical error being less than 1\%. The total $p$ flux error 
is 4\% at 1 GV, <2\% between 2 and 350 GV, and  6\% above 1.1 TV.
\begin{figure}
\begin{center}
\vspace{2mm}
\subfigure[]
{
\includegraphics[height=6.5cm, width=5.4cm,angle=270]{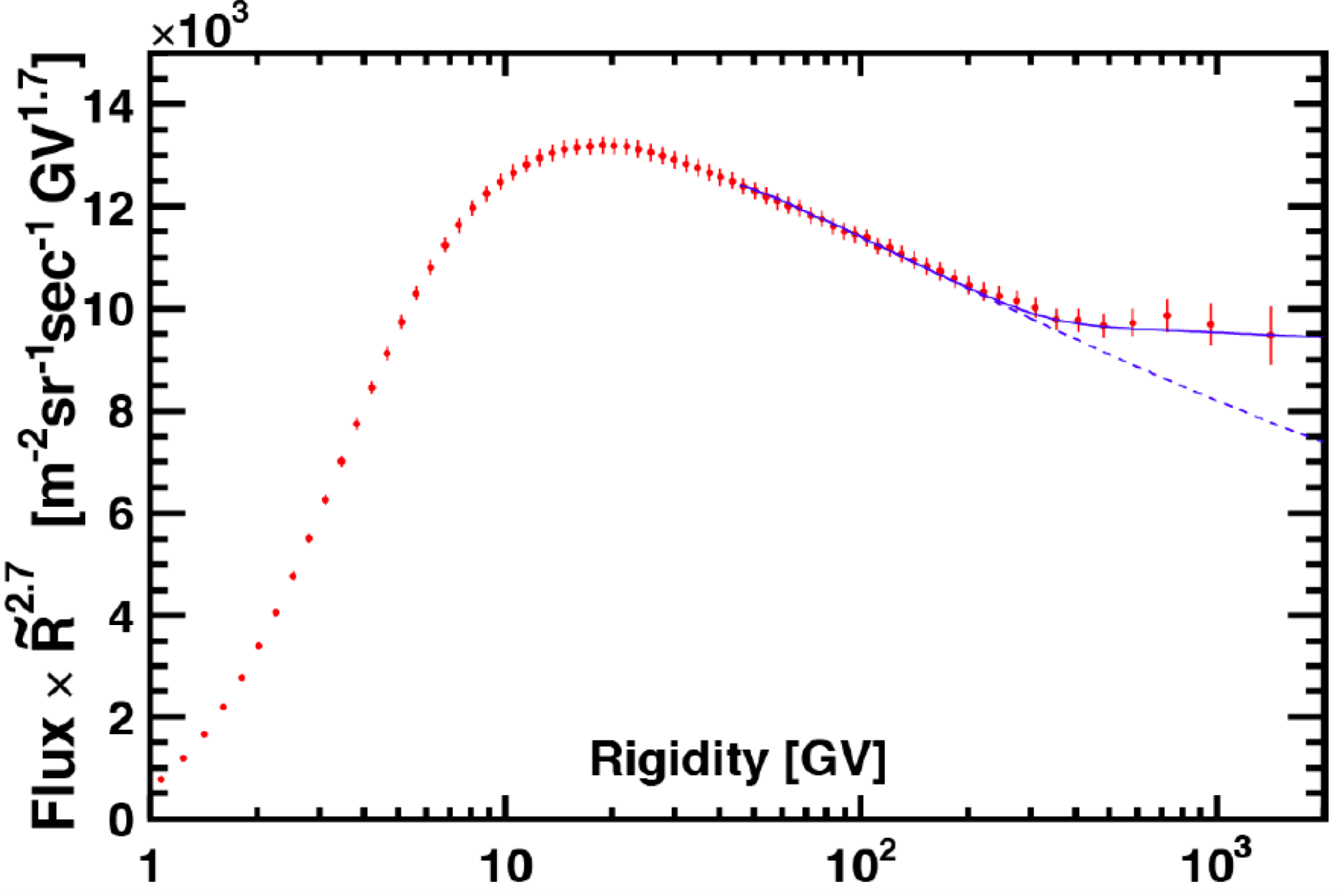}
\label{pfluxa}                            
}
\hspace{10mm}
\subfigure[]
{
\includegraphics[height=6.5cm, width=5.4cm,angle=270]{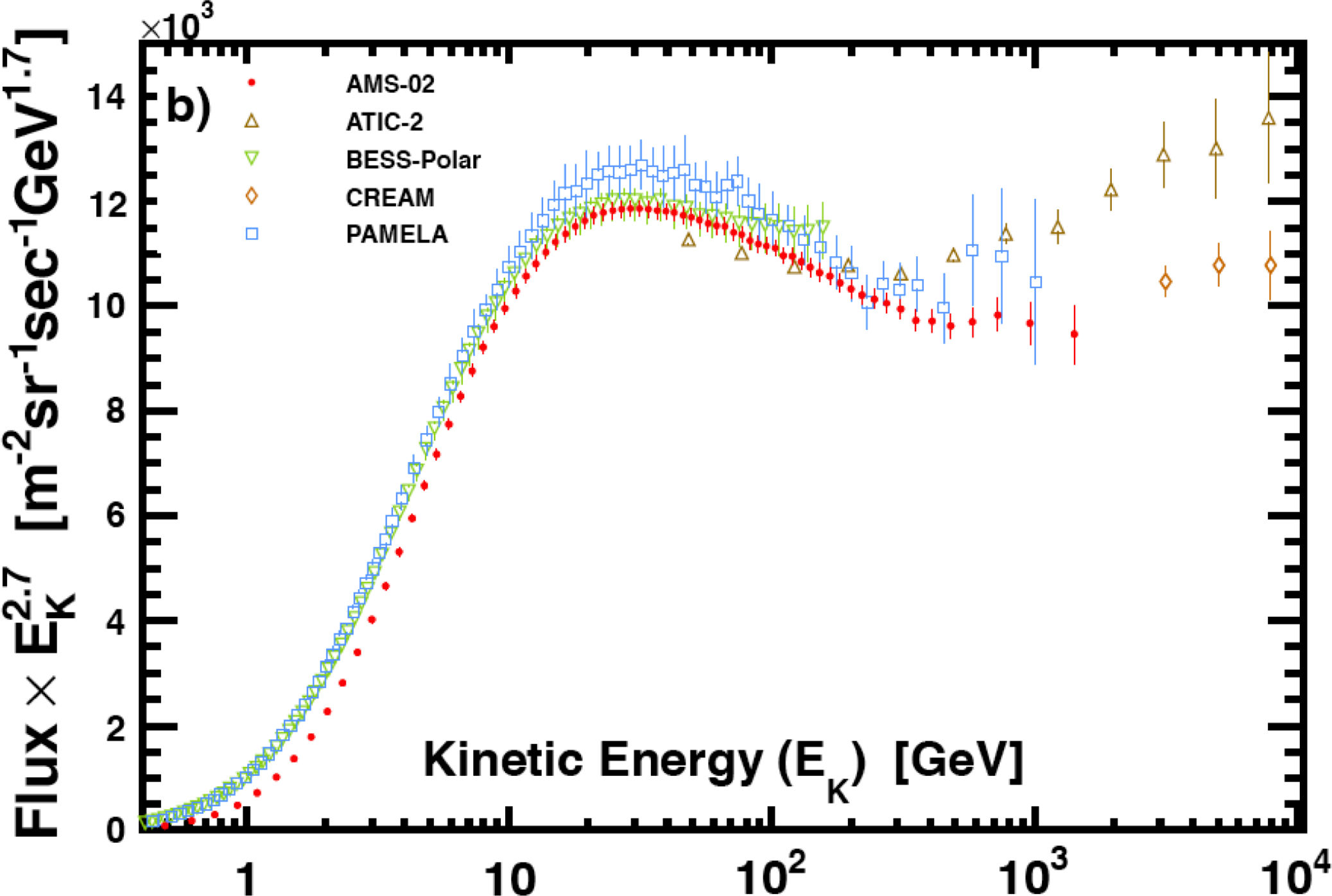}
\label{pfluxb}                            
}\\
\vspace{2mm}
\subfigure[]
{
\includegraphics[height=6.5cm, width=5.4cm,angle=270]{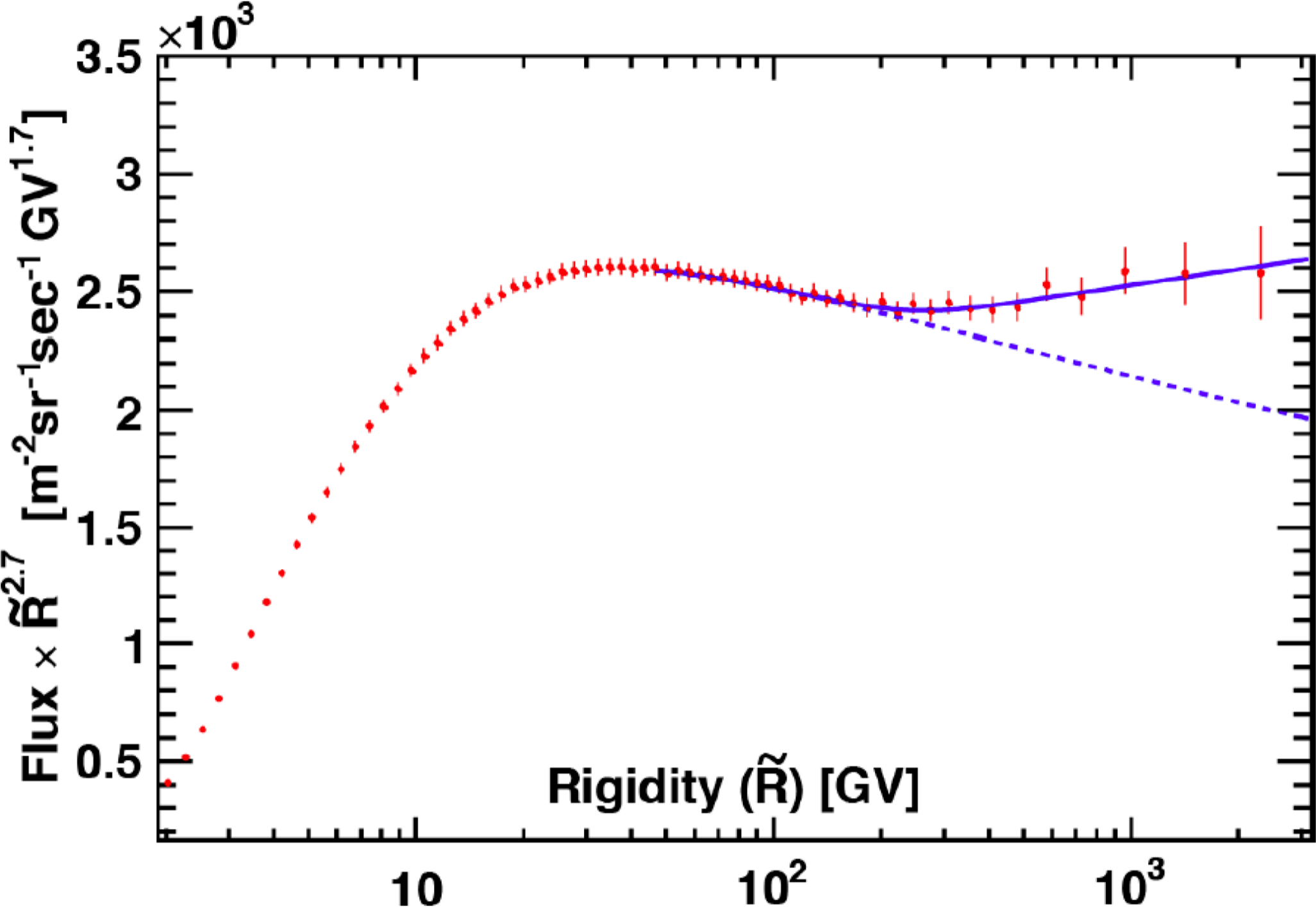}
\label{Hefluxa}                            
}
\hspace{10mm}
\subfigure[]
{
\includegraphics[height=6.5cm, width=5.4cm,angle=270]{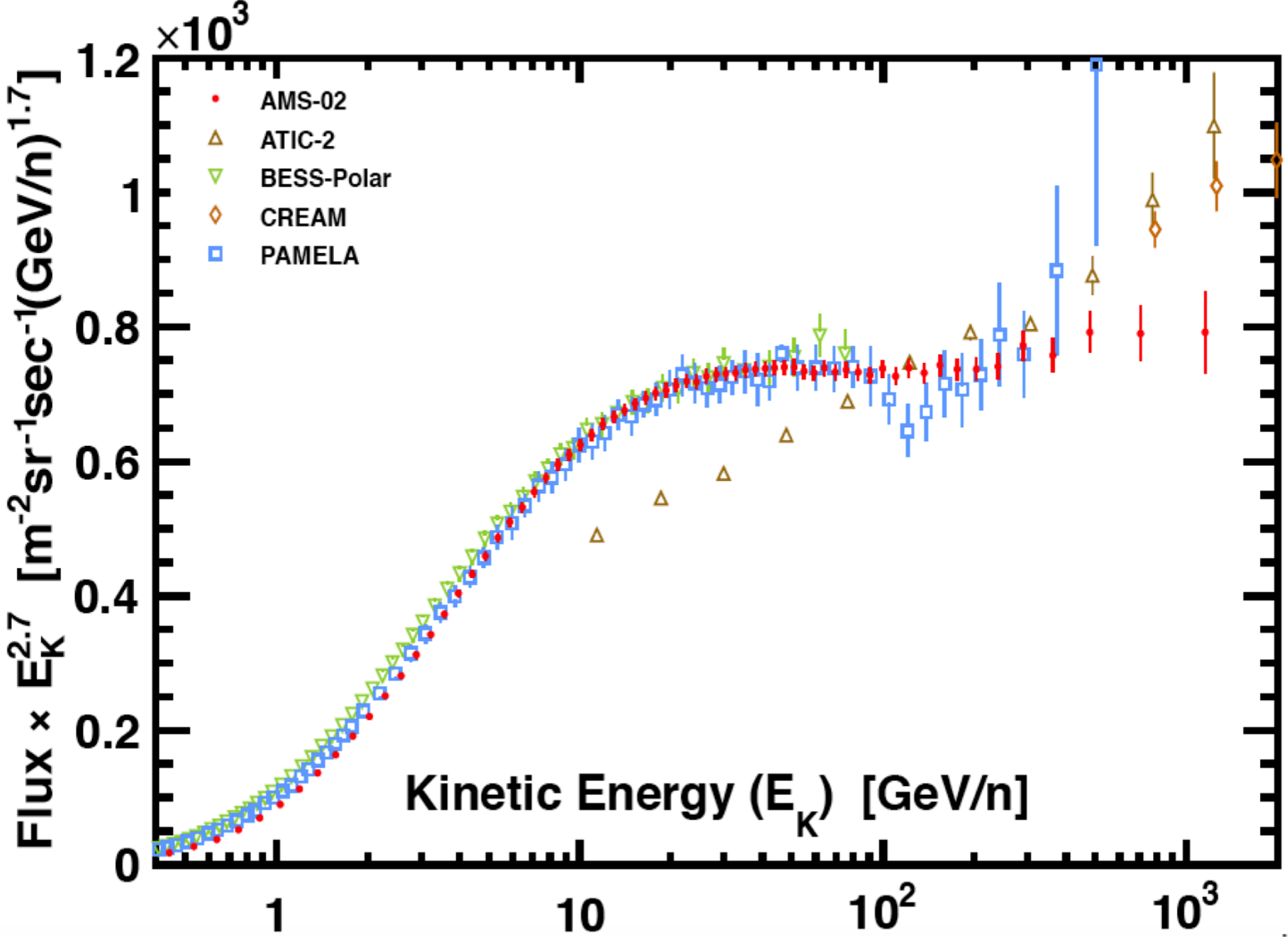}
\label{Hefluxb}                            
}
\caption{AMS (a) $p$ and (c) He  fluxes multiplied by $R^{2.7}$ as a function of rigidity $R$. Curves represent the fit of
Eq.~\ref{eq1}  (solid line) and a single power-law  (dotted line) to the data.
AMS (b) $p$ and (d) He fluxes multiplied by $E_k^{2.7}$  as a function of kinetic energy per nucleon  $E_k$, compared with the most recent
results from spectrometers (BESS, PAMELA) and calorimetric experiments (ATIC, CREAM)\cite{AMS-proton, AMS-helium}.
}
\label{pflux}
\vspace{-4mm}
\end{center}
\end{figure}
The AMS $p$ and He spectra are shown in Figs.~\ref{pfluxa} and \ref{Hefluxa}, respectively.
They clearly show a deviation with respect to a single power-law behaviour $R^\gamma$ around 300 GV. 
The data above 45 GV can be fitted  to  a double power-law function
\begin{equation}
\Phi = C \left(\frac{R}{45\, GV}\right)^{\gamma} \left[ 1 + \left( \frac{R}{R_0}\right)^{\frac{\Delta\gamma}{s}} \right]^s
\label{eq1}
\end{equation}
with a smooth transition from a spectral index $\gamma$ below the transition rigidity $R_0$
to $\gamma+\Delta \gamma$ above. 
The fitted spectral hardening is 
$\Delta\gamma$= 0.133$^{+0.032}_{-0.021}$(fit)$^{+0.046}_{-0.030}$(sys)$\pm$0.005(sol)
\footnote{The first error (fit) stems from statistical and uncorrelated systematic errors 
of the flux. The second (sys) is the error from the remaining systematic errors; the third (sol) is due
to the variation of the solar potential \cite{AMS-protonPRL}. 
}
for $p$ occurring at $R_0$= 336$^{+68}_{-44}$(fit)$^{+66}_{-28}$(sys)$\pm$1(sol) GV. 
For He, $\Delta\gamma$= 0.119$^{+0.013}_{-0.010}$(fit)$^{+0.033}_{-0.028}$(sys)
and $R_0$= 245$^{+35}_{-31}$(fit)$^{+33}_{-30}$(sys) GV.
Then $p$ and He spectra have very similar rigidity dependence, 
but the He spectral index is different from that of $p$: $\gamma_{He}=$-2.780$\pm$0.005(fit)$\pm$0.001(sys)
vs. $\gamma_{p} =$-2.849$\pm$0.002 (fit)$\pm^{+0.004}_{-0.003}$(sys) $^{+0.004}_{-0.003}$(sol), respectively.
Moreover, by studying the detailed variation of $\gamma$ with rigidity, it turns out  that 
the spectral index  progressively hardens above 100 GV for both  species, 
but still $p$ is softer than He.  
The $p$/He flux ratio can be fitted to a single-power law in rigidity 
above 45 GV with a constant spectral index -0.077$\pm$0.002 (stat)$\pm$0.007 (sys) \cite{AMS-helium}.\\
In Figs.~\ref{pfluxb} and \ref{Hefluxb} the AMS results are compared with the most recent 
measurements by magnetic spectrometers (PAMELA \cite{PAMELApHe}, BESS-Polar II \cite{Bess-pHe}) and balloon-borne 
calorimetric experiments (ATIC \cite{ATIC}, CREAM \cite{CREAMpHe}). 
In \cite{PAMELAhl}, it was pointed out that PAMELA and AMS results agree very well: 
in the energy region not affected by solar modulation, i.e. at rigidities $>$30 GV, the average  
ratio between the fluxes measured by the two experiments is 0.988 for $p$ and 1.036 for He. 
It is the first time in the history of CRs that the absolute fluxes reported by different experiments differ at the level of percent.
Another remark  is that, 
unless exploiting the transition radiation detector to extend the range of energy measurement for 
$Z\ge2$ nuclei, as proposed in \cite{Obermeier}, AMS cannot push the measurement of
$p$ and He spectra beyond its MDR, which is few TeV/n. Then next
accurate measurements of $p$ and He fluxes over the spectral break and extending up to hundreds of TeV
might come from future calorimetric experiments (see section \ref{s:futurexp}).

These new results force us to revise the standard model of GCRs 
and find theoretical explanations for the origin of the spectral hardening.
The standard paradigm is 
based on diffusive shock acceleration (DSA) of charged particles 
in supernova remnants (SNRs), with an injection spectrum described by a power-law in energy $E^{\alpha}$,
followed by CR propagation in the ISM, with an escape time from the Galaxy proportional to $E^{\delta}$.
It predicts that  the spectra of primary CRs observed at Earth follow
a power-law  $E^{\gamma}$  with 
spectral index $\gamma=\alpha+\delta\approx$-2.7, i.e.  steeper than the injection spectra.
Several theoretical break models have been proposed to explain the observed deviation from a single power-law behaviour; 
they can be grouped essentially in three main classes \cite{Serpico}. 
In the first class, the hardening is interpreted as a feature due to the
acceleration mechanism of CRs at the source.
For instance, it could be the effect of distributed acceleration by
multiple SNRs in OB associations or Superbubbles \cite{Butt, Parizot},  
or an effect of  reacceleration of CRs by weak shocks in the Galaxy  \cite{Thoudam}, 
or alternatively it could  represent the spectral concavity caused by the
interactions of accelerated CRs with the accelerating shock, as foreseen 
in non-liner DSA models \cite{DSA,DSA2}.
In the second class of theoretical break models, 
the spectral hardening could originate from 
subtle effects of CR propagation, like 
an inhomogeneous diffusion, characterized by 
a different energy dependence  in different region of the Galaxy \cite{Tomassetti1}, 
or  a non-linear coupling of CRs with the diffusion coefficient, typical of 
models in which the transport of GCRs is regulated by self-generated waves \cite{Blasi0, Nava}.
Finally, the spectral hardening  could  be due to the effect of 
local young sources \cite{Thoudam2, Liu, Zhang},
or to a mixture of fluxes accelerated by 
sources with different properties (e.g. few old local SNRs vs. an ensemble of young and distant SNRs) 
contributing in different energy region of the spectra \cite{Tomassetti2}.
Different ideas were also explored to explain the softer $p$ spectrum with respect to 
He and heavier nuclei, e.g.
different astrophysical sources for $p$ and He \cite{Zatsepin},
an injection in the acceleration cycle more efficient for He than for $p$ \cite{Malkov}, 
variable He and $p$ concentrations in SNR environments \cite{Drury,Ohira},
different spallation rates for $p$ and heavier nuclei during propagation time \cite{Blasi}. 
\begin{figure}[!h]
\vspace{6mm}  
\begin{center}                                 
\includegraphics[width=.7\textwidth,angle=0]{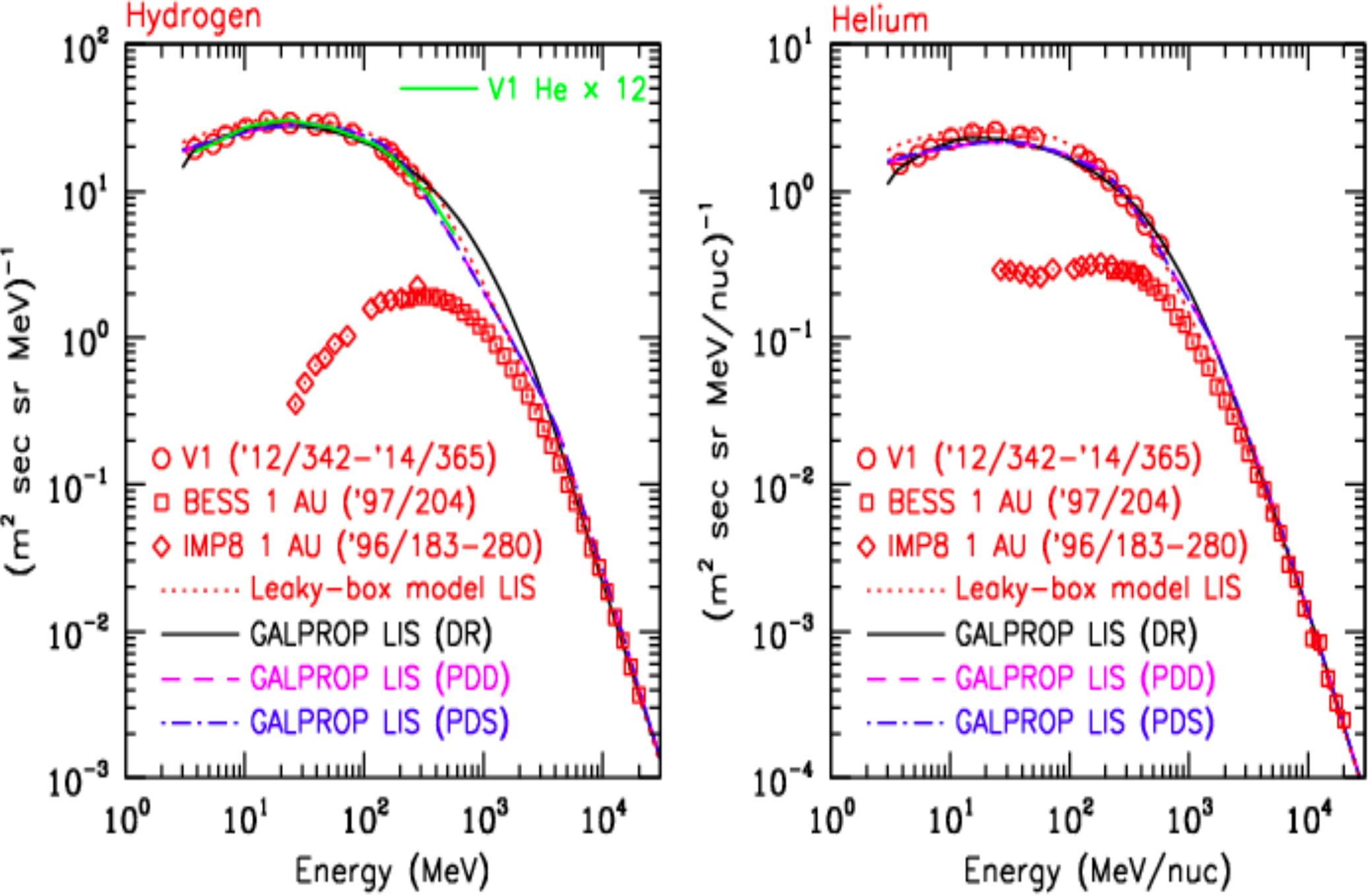}
\vspace{3mm}  
\caption{Energy spectra of H (left) and He (right) from Voyager 1, compared with solar-modulated spectra at 1
AU from  BESS and IMP8, and predictions from different propagation models \cite{Cummings}.
}
\label{HHeV1}                                 
\end{center}
\end{figure}
\subsection{Low-energy region (<10 GeV)}
The large statistics of low-energy proton data collected by PAMELA and AMS 
is extremely important to enhance our understanding of solar modulation and geomagnetic effects on CRs. 
The first can be investigated by studying the time evolution of proton flux over several years.
At the ICRC, AMS reported the detailed time variation of the $p$ flux from 1 to 10 GV over the period 2011-2013, 
showing not only  the expected overall flux decrease with increasing solar activity until its maximum in 2014, but also  
periodic monthly variations of the flux related to strong solar events, i.e. Coronal Mass Ejections and Forbush decreases
\cite{AMS-lowenergy}. 
New accurate measurements of under-cutoff proton fluxes from 70 MeV up to few GeV at low Earth orbit
were presented by PAMELA. Protons were classified into geomagnetically trapped and re-entrant albedo on the basis of particle tracing techniques
in a wide geomagnetic region, allowing more accurate view of atmospheric and geomagnetic effects on the CR transport, 
and a better characterization of the radiation environment near the Earth \cite{PAMELA-lowenergy}.

The spectra of H and He were measured for the first time in their unmodulated state by 
the Cosmic Ray Subsystem (CRS) experiment on Voyager 1 (V1).
The V1 space probe left the heliosphere on August 2012 and entered the denser and colder environment of the interstellar cloud surrounding
the solar system.  After crossing the heliopause at a radial distance of 121.6 AU, 
the intensity of GCRs reached the highest level observed over the 35 years of V1 journey and remained constant for the next two years. 
At the same time a depletion of heliospheric ions, which escaped into the interstellar space, was observed.
It turned out that the heliosphere operates as a shield that excludes more than 75\% of the GCRs with energy $>$ 70 MeV \cite{Stone}.
Then the data collected in the ISM by the CRS solid state detector telescopes
allowed to reveal the low energy part of the interstellar CR spectra from H to Ni, 
i.e. not modulated by the time-dependent solar wind.   
Very similar spectra for H and He  have been measured in the range 3-346 MeV/n, as shown in Fig.~\ref{HHeV1}.
The H/He ratio is $\sim$12 independent of energy. Both the H and He spectra exhibit a broad peak 
at the same energy per nucleon, 
implying that thay are not affected by solar modulation and supporting the idea that V1 is really travelling in the local ISM.
The roll over at low energies from a power-law at high energies
is likely a consequence primarily of ionization energy losses  and suggests that V1 is not in
the nearby  of a recent source of GCRs \cite{Cummings}.

\begin{figure}[h]
\begin{center}
\subfigure[]
{
\includegraphics[height=6.6cm, width=5.4cm,angle=270]{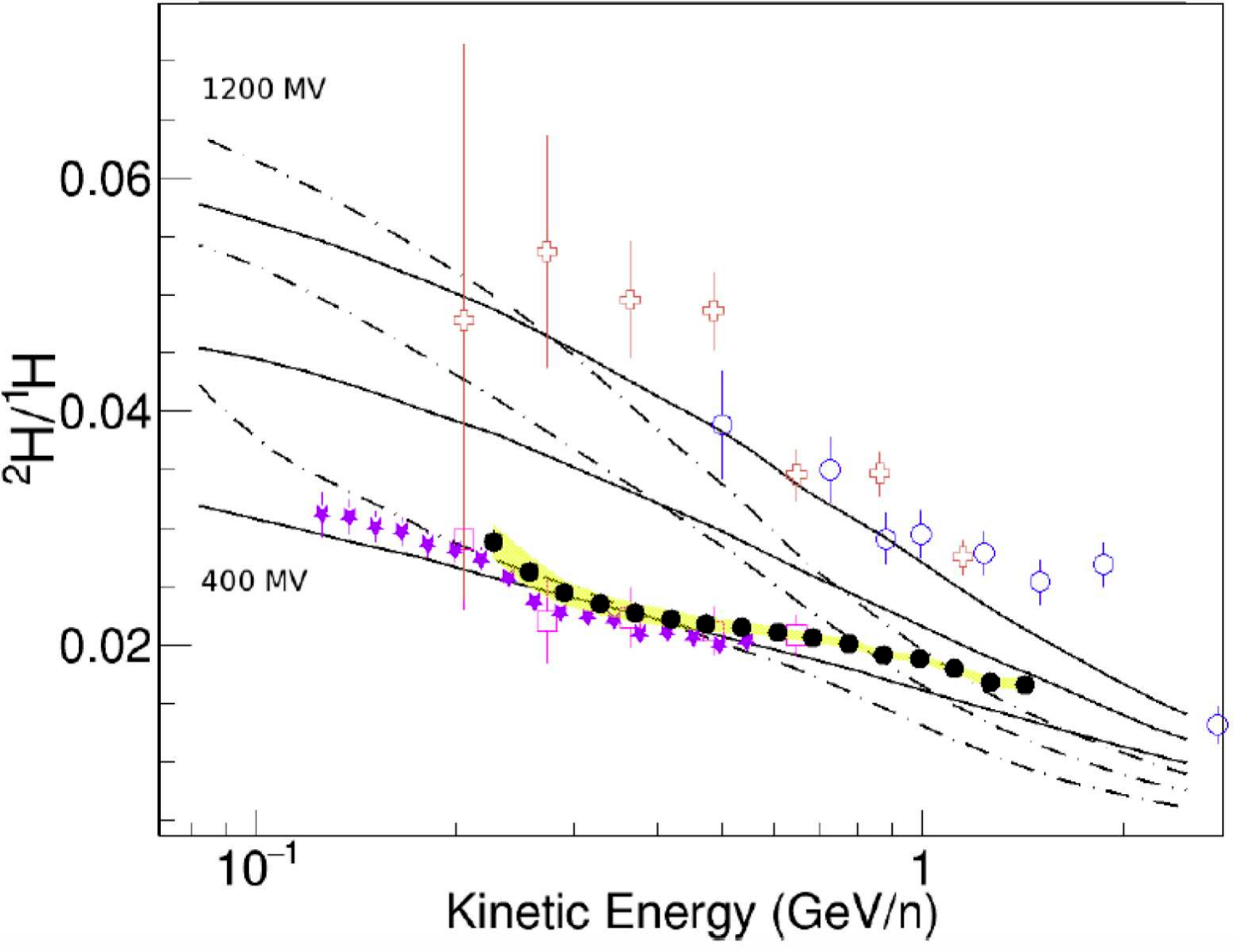}
\label{BESS_H}                            
}
\hspace{7mm}
\subfigure[]
{
\includegraphics[height=6.6cm, width=5.4cm,angle=270]{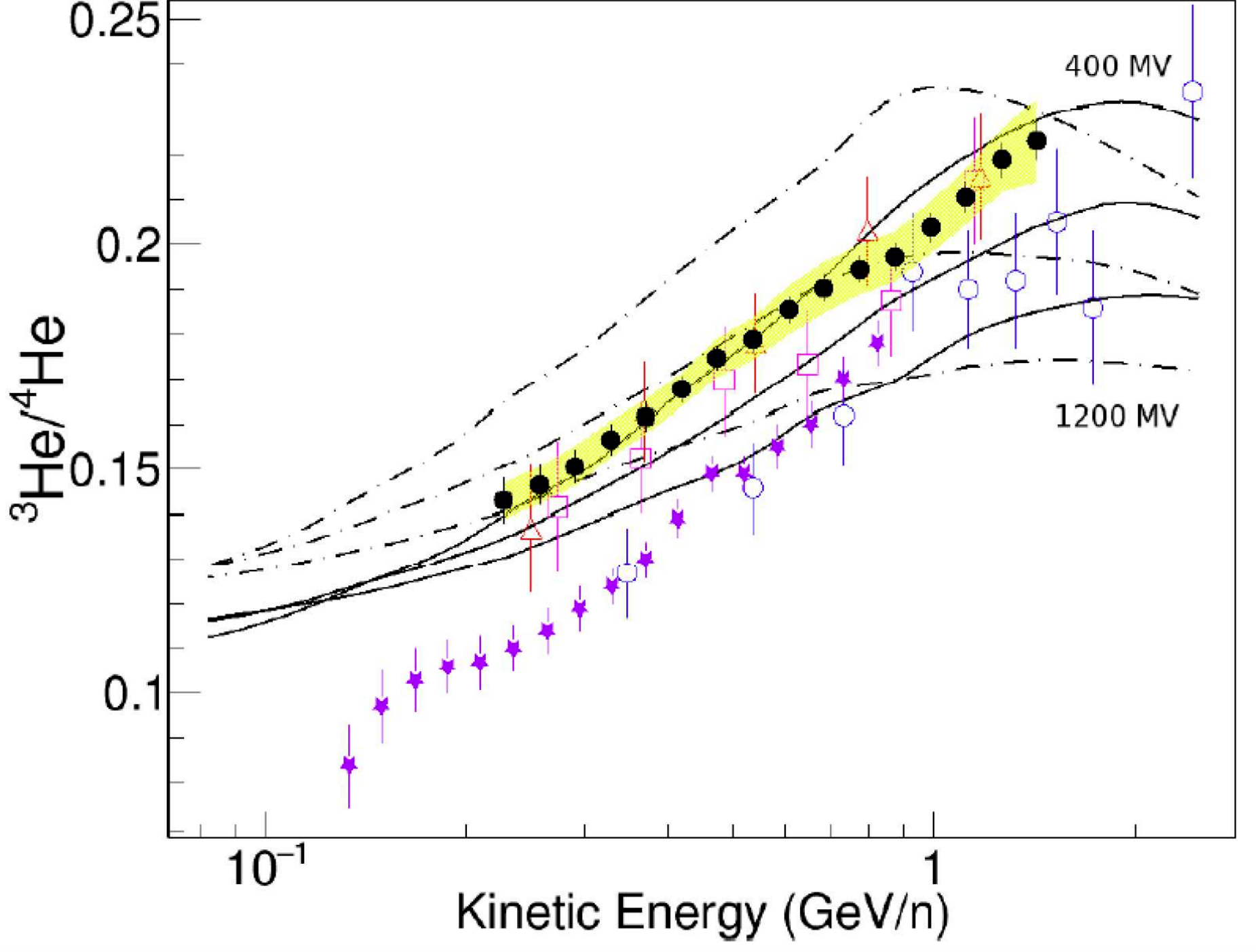}
\label{BESS_He}                            
}
\caption{Ratios of (a) $^2$H/$^1$H and (b) $^3$He/$^4$He as measured by BESS-Polar II (filled circles), 
compared with previous data from IMAX-92  (open circles), BESS-93 (open squares),
AMS-01 (open triangles) and PAMELA (filled stars).
The lines represent ``Plain Diffusion'' (solid) and ``Reacceleration'' (dashed) models, calculated with GALPROP
for different solar modulation parameters  \cite{Bess-HHeIsot}.
}
\label{BESS}
\vspace{-0.5cm} 
\end{center}
\end{figure}

New isotopic measurements of H and He were also presented  by PAMELA and BESS-Polar.
$^2$H and $^3$He are mostly secondary particles produced by the interaction of primary $^1$H and $^4$He CRs with the ISM
 during propagation. Then the secondary-to-primary ratios $^2$H/$^1$H and  $^3$He/$^4$He
 provide  important information on propagation history of CRs in the ISM, 
and they are complementary to B/C measurements in constraining propagation models.  
Combining a tracking system based on a 0.8 T superconducting magnet and a JET-type drift chamber,
with a time-of-flight hodoscope and an aerogel Cherenkov counter, the balloon-borne BESS-Polar II instrument
measured these isotopes with  an excellent mass separation 
up to 1.5 GeV/n.
The isotope ratios from data taken in a 24.5-day Antarctic flight  in 2007, during a solar minimum, 
are shown in Fig.~\ref{BESS}.
These results have unprecedented precision and allow to discriminate between different propagation models \cite{Bess-HHeIsot}.
While the H isotopes ratio is consistent with previously published PAMELA data \cite{Pamela-HHeIsotOLD}, 
taken during the same solar minimum, 
there is an evident disagreement as far as $^3$He/$^4$He ratio is concerned. 
PAMELA presented new preliminary measurements of the isotopic composition \cite{Pamela-HHeIsot},
based on a more refined analysis procedure than the previous work \cite{Pamela-HHeIsotOLD} and extending to higher energies, 
but a comparison with the new BESS-Polar data was not yet shown. 
\section{Nuclei with Z$>$2}
\label{s:nuclei}
\subsection{Light nuclei}
In the last decade the balloon-borne experiments CREAM \cite{CREAM-heavy} and TRACER \cite{TRACER}
measured the energy spectra of the 
most abundant primary heavy nuclei (C, O, Ne, Mg, Si, Fe) up to energy of $\sim$10$^{14}$ eV per particle.
All the spectra are well fitted to  single
power-laws in energy with a remarkably similar spectral index
$\gamma\approx$-2.65 which well agrees with that of He in the multi-TeV region, 
implying that all these elements likely have the same origin
and share the same acceleration and propagation processes.
However, by fitting simultaneously the normalized energy spectra to a broken-power law, 
CREAM reported also a hint of a spectral hardening above $\sim$200 GeV/n \cite{CREAM-harden}, 
similar to that observed for $p$ and He.
At the moment, the statistical and systematic uncertainties of
these measurements significantly hinder a conclusive interpretation of the data.
Further accurate observations are needed 
to extend with high statistics the present data to higher energies, on the one hand, 
and measure accurately a possible curvature of the spectrum and the position of the spectral break-point for individual nuclear species, 
on the other hand.
While only future calorimetric experiments will be able to address the first task, 
AMS has the unique opportunity to accurately investigate the 
region between 100 GeV/n and few TeV/n to search for possible spectral features. \\
\begin{figure}
\begin{center}
\subfigure[]
{
\includegraphics[height=6.6cm, width=5.4cm,angle=270]{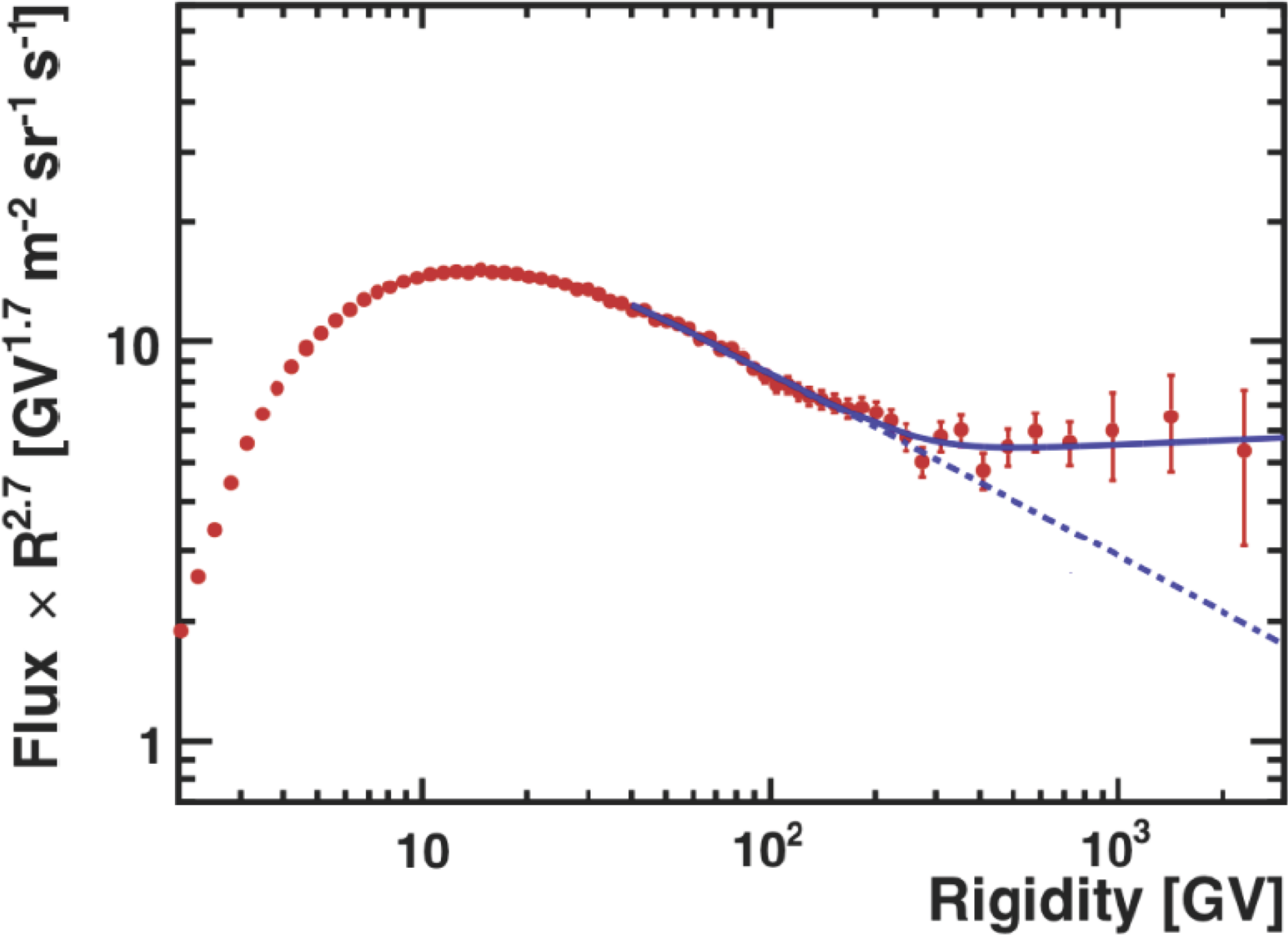}
\label{Liflux}                            
}
\hspace{10mm}
\subfigure[]
{
\includegraphics[height=6.6cm, width=5.4cm,angle=270]{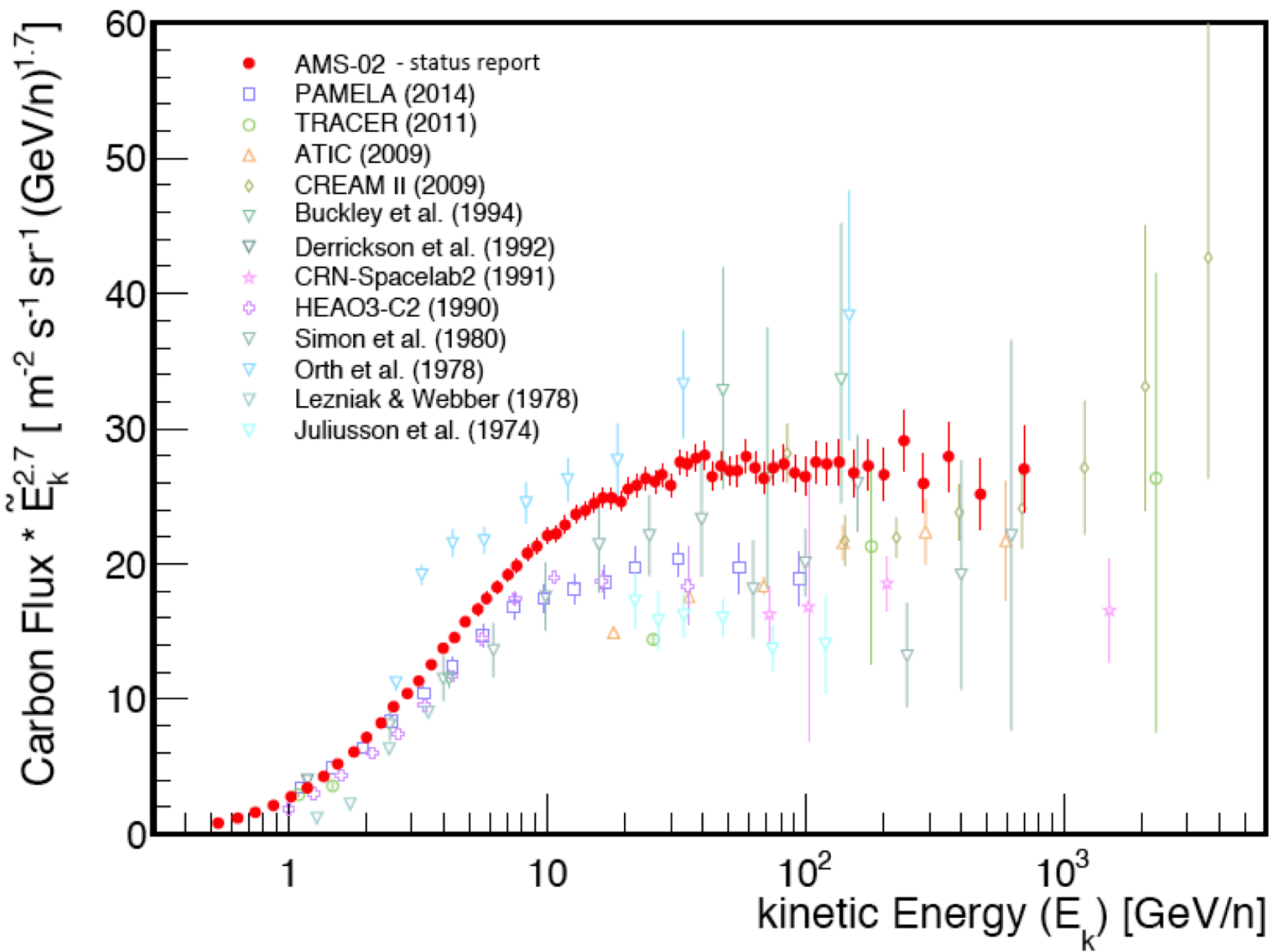}
\label{Cflux}                            
}
\caption{(a) The AMS Li flux multiplied by $R^{2.7}$ as a function of rigidity $R$  \cite{AMS-Li}. 
(b) The AMS C flux multiplied by $E_k^{2.7}$ as a function of kinetic energy per nucleon $E_k$, 
compared with earlier results \cite{AMS-C}. 
}
\label{LiCflux}
\vspace{-0.4cm} 
\end{center}
\end{figure}
So far, AMS reported only preliminary measurements of the energy spectra of lithium and carbon.
Lithium, like Be and B, has a  secondary origin, i.e. it is produced by the spallation of heavier 
CR nuclei with the ISM during their propagation. 
Measurements of these light nuclei spectra, as well as of secondary-to-primary abundance ratios,  
allow to probe galactic propagation models and  constrain their parameters.
Since Li accounts for only about 0.1\% of CRs, 
its measurement is really challenging and few data  
from old balloon experiments were available so far. 
Thanks to its large geometrical factor and exposure, 
AMS measured about 1.5$\times$10$^6$ Li events. The
 independent charge identification provided by TOF and tracker
allowed to select a pure sample of Li events, with small  contamination of  
heavier CR nuclei interacting in the upper part of the detector
and producing lithium.
The differential Li flux has been measured between 2 GV and 3 TV \cite{AMS-Li}, 
showing an unexpected  hardening above 200 GV (Fig.~\ref{Liflux}), the same rigidity region as for $p$ and He.
Surprisingly such a feature is not observed in the AMS carbon spectrum \cite{AMS-C}, 
which above 45 GV seems to be compatible with a single power-law (Fig.~\ref{Cflux}).
Nevertheless, it has been pointed out that, unlike the $p$ and He data,
these AMS measurements are preliminary, since 
the statistical errors are still dominant  over systematics above 100 GV; 
more data need to be collected to accurately determine the shape of these spectra  
at high energies. 
It is also worth  noting the significant discrepancy between 
 the AMS and PAMELA carbon fluxes,
which clashes with the general good agreement observed between their $p$ and He data. 
New preliminary results on the Li and Be fluxes from 1 to 100 GV \cite{PAMELA-LiBe}
and their isotopic composition up to 1.2 GeV/n \cite{PAMELA-LiBeIsot}
were also presented at the ICRC by PAMELA, but these analyses are still in an early stage. \\
\begin{figure}
\begin{center}                                 
\includegraphics[height=7.6cm, width=5.4cm,angle=270]{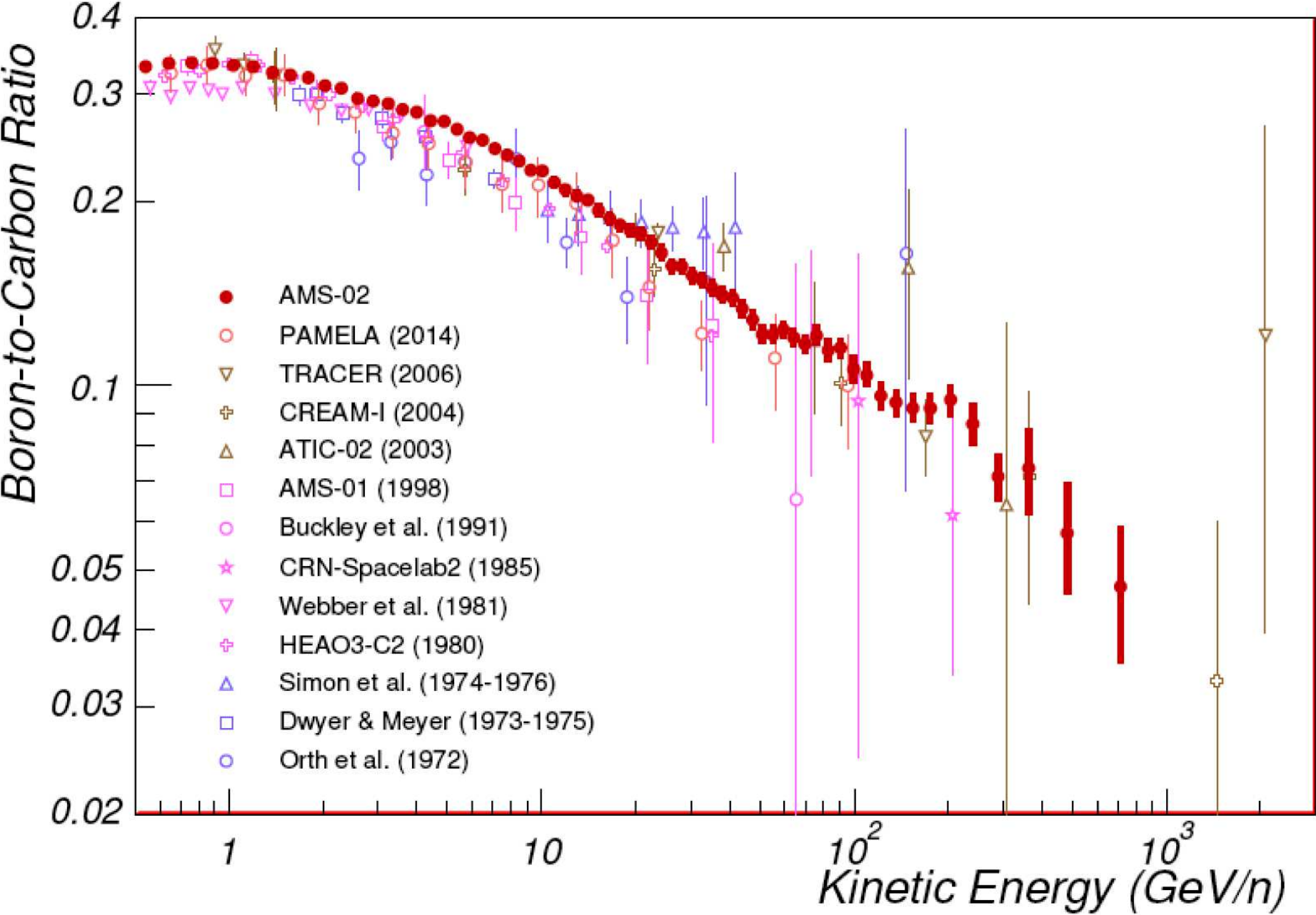}
\caption{New AMS B/C ratio compared with a compilation of earlier measurements \cite{AMS-BC}. 
}
\label{BC}   
\vspace{-0.8cm}                                                   
\end{center}
\end{figure}
Information about the diffusion of CRs in the
Galaxy can be inferred from the measurement of the secondary-to-primary abundances (e.g. B/C, sub-Fe/Fe
ratios), which allow to estimate the average amount of material that CRs typically traverse between injection
and observation. The escape path-length $\lambda$ of CRs from the Galaxy is known to decrease as the particle rigidity increases, 
following a power-law $\lambda\propto R^{\delta}$, where $\delta$ is the diffusion coefficient index. 
This is a key parameter and its accurate measurement is crucial 
to derive the injection spectrum at the source by correcting the observed
spectral shape for the energy dependence of the propagation term.
Pre-AMS B/C data 
favor a value of $\delta\approx$-0.6 at  energies of few GeV/n, 
while  high-energy data 
seems to flatten to $\delta\approx$-0.4.
However measurements  by balloon-experiments (CREAM \cite{CREAM-BC}, ATIC \cite{ATIC-BC}, TRACER \cite{TRACER-BC}) 
between 10$^2$ and 10$^3$ GeV/n, 
suffer from statistical limitations and large systematic errors stemming from the corrections for
nuclei produced by CR interaction with the residual atmosphere, and do not
allow to tightly constrain  the value of $\delta$ and discriminate among different propagation models. 
The new AMS data (Fig.~\ref{BC}), based on 7$\times$10$^6$  C and 2$\times$10$^6$  B events, 
have unprecedented precision and accurately  determine 
the spectral shape of the B/C ratio up to about 700 GeV/n \cite{AMS-BC}.
Given this high level of precision,  the theoretical uncertainties 
affecting the models become the major limitation
for the interpretation of data. In \cite{Genolini},
it is shown that the lack of information about a  possible injection of B nuclei at the sources 
and the errors on the nuclear cross sections
seriously limit the extraction of CR diffusion parameters from B/C data.
In \cite{Kunz}, the transport parameter space is scanned 
to find solutions for the propagation equations fitting simultaneously pre-AMS B/C and other CR data.
A large number of degenerated solutions were found which 
cannot be even completely resolved by further constraints from AMS data, but 
might be only reduced with a multi-messager approach. \\
Accurate measurements of B/C ratio at TV rigidities are needed to test the spectral break models. 
It has been pointed out that an error <10\% on the B/C ratio at 1 TeV/n would allow 
to discriminate between different classes of models \cite{Kunz}.
AMS can reach this goal 
by collecting more data in the next years, since its preliminary result
is still dominated by the statistical uncertainties above 60 GV.
A propagation induced break would imply secondary spectra with a more pronounced break than primary ones, 
while no feature are expected in secondaries/primaries ratio if the break origin is in the source \cite{Serpico}.
The significant spectral break observed by AMS for Li seems to point to the first scenario, 
but more data on other nuclei are needed to clarify the picture.
However, in \cite{Aloisio}, the authors observe that  B/C ratio above 100 GeV/n is somewhat higher than the prediction 
of a non-linear CR Galactic transport model used to explain the spectral break of $p$ and He fluxes.
That could indicate a contribution of secondary production at the source that may become
significant at TeV/n scale, thus further complicating the possibility to disentangle
source and propagation effects.
\subsection{Iron and ultra-heavy nuclei}
Elements above Fe are extremely rare compared to light elements,
but they can provide unique information about the site of acceleration of GCRs. 
Measurements of the abundances of elements from Z=30 to Z=38 by the Trans-Iron Galactic Element Recorder (TIGER) \cite{TIGER}
and the  ratios of $^{22}$Ne/$^{20}$Ne, $^{58}$Fe/$^{56}$Fe, and C/O  \cite{ACECRISOLD}
by the Cosmic Ray Isotope Spectrometer (CRIS) onboard the Advanced Composition Explorer (ACE), 
support the idea of OB associations as CR acceleration sites.
Since supernovae (SN) explode  preferentially in groups of massive stars, named OB associations,
then DSA would occur in an ISM of solar-system composition enriched in freshly synthesized material from previous SN
or Wolf-Rayet star ejecta.
\begin{figure}
\begin{center}
\vspace{1mm}
\subfigure[]
{
\includegraphics[height=6.2cm, width=5.4cm,angle=270]{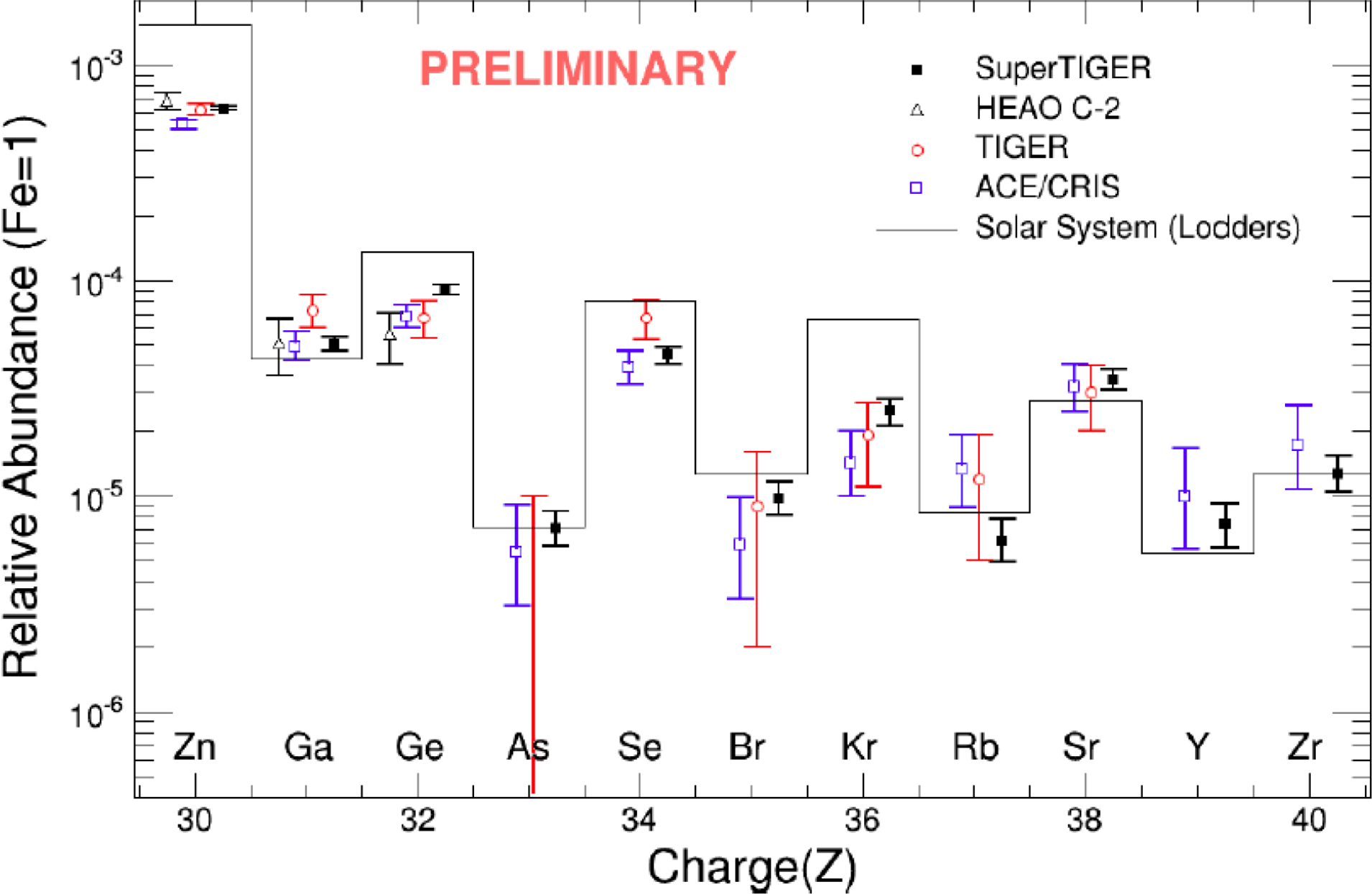}
\label{UHGCRa}                            
}
\hspace{5mm}
\subfigure[]
{
\includegraphics[scale=0.38,angle=270]{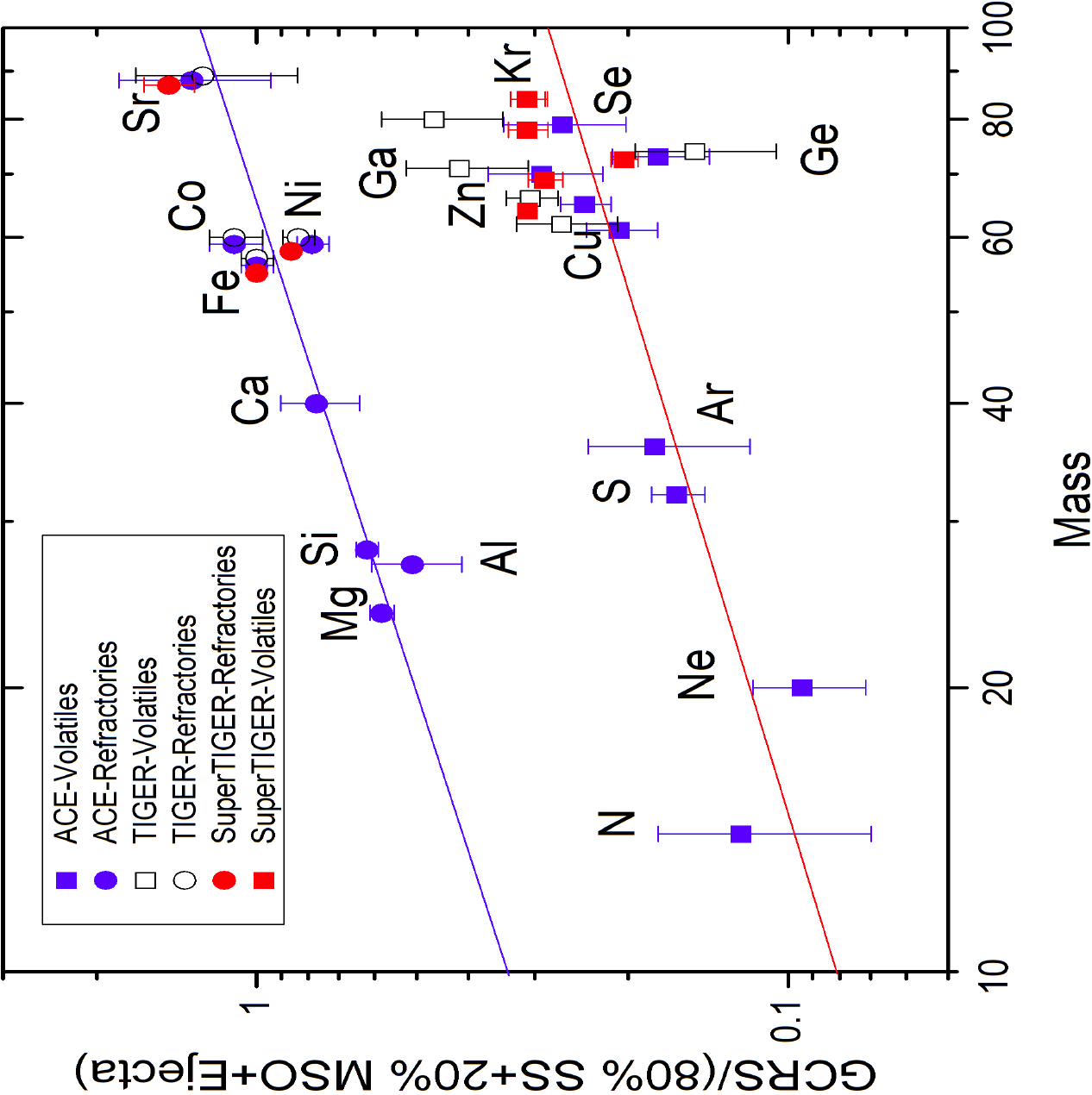}
\label{UHGCRb}                            
}
\caption{(a) Comparison of SuperTIGER Top-of-Atmosphere abundances with TIGER data and  space abundances
from ACE-CRIS, and HEAO.
(b) GCR source  abundances compared to a mixture of 80\% solar system 
and 20\% from massive star outflow as a function of the nuclear mass.
Preliminary SuperTIGER data (the errors are  only statistical)
are compared with previous results from ACE-CRIS and TIGER \cite{STIGER2}.}
\label{UHGCR}
\vspace{-0.8cm}
\end{center}
\end{figure}
New data confirming this scenario have been presented by the SuperTIGER collaboration. 
Featuring a 8.3 m$^2$sr acceptance, this balloon-borne experiment measured 
the abundances of ultra-heavy GCRs (UHGCRs) from Z=30 up to Z=40
with unprecedented statistics 
in a 55-day flight over Antarctica in 2012-13 \cite{STIGER}. 
The SuperTIGER instrument is made of two identical modules, each one consisting
of two scintillating fiber hodoscopes for particle tracking, three scintillators and two Cerenkov detectors, 
equipped with acrylic and aerogel radiators respectively, to measure the particle charge and velocity in the range  0.8-10 GeV/n.
The instrument is capable of separating each individual species from
Ne to Ba with excellent resolution (0.16 charge units for Fe) \cite{STIGER1}. 
Observed elemental abundances of UHGCRs are 
corrected first for the residual atmospheric overburden (Fig.~\ref{UHGCRa}) and
 then  for the effects of 
fragmentation in the ISM to derive the source abundances. 
The ratios of these calculated GCR source abundances to a  mixture of 
$\sim$20\%  ejecta of massive stars (including Wolf-Rayet stars and core-collapse supernovae) mixed 
with $\sim$80\% material of solar system composition, are plotted as a function of the elemental atomic mass in Fig.~\ref{UHGCRb}.
By assuming this source mixture, a clear separation of refractory and volatile elements and a better ordering  by mass is obtained
with respect to a pure solar system composition, indicating that 
the refractory elements (found in interstellar dust grains) are more effectively accelerated than volatile ones (present in interstellar gas) \cite{STIGER2}.
These preliminary  measurements 
are consistent with previous
observations by TIGER and ACE-CRIS (performed in a lower energy range, 150-600 MeV/n), 
and provide an improved test of the origin of GCR in OB associations and the volatility model of GCR acceleration.\\
SuperTIGER also presented a very preliminary measurement of (Sc+Ti+V)/Fe ratio 
in two energy bins defined by the thresholds of the two Cerenkov detectors \cite{STIGER3}. 
Future plans are to calibrate their velocity scale and  study the energy dependence 
of this secondary-to-primary ratio with finer resolution. 

Another proof supporting the OB associations model for GCR origin, 
comes from the first measurement of a primary cosmic-ray clock, reported by ACE-CRIS \cite{ACECRIS}.
CRIS can provide an excellent elemental and isotopic separation by
measuring the dE/dx and total energy of CRs stopping in a stack of silicon detectors. 
Based on 16.8 years of collected data, 
CRIS was able to detect 15 well-resolved ($\sim$0.24 mass resolution) 
events of the rare radioactive $^{60}$Fe isotope 
in a set of  2.95$\times$10$^5$  $^{56}$Fe nuclei, measured
in the 240-470 MeV/n energy interval.
A detailed analysis demonstrated that at most 2 out of 15 $^{60}$Fe could have been produced 
by interstellar fragmentation of Ni isotopes or 
could be spill-over from $^{58}$Fe.
Then  the observed $^{60}$Fe  are almost all primary nuclei. 
Since $^{60}$Fe decays by $\beta^-$ with a half-life of 2.62 Myr, 
the fact that we observe this radioisotope near Earth 
implies that the time $T$ elapsing between nucleosynthesis and 
CR acceleration is $\le$10 Myr.
Moreover, a lower limit  of 10$^5$ yr is inferred for $T$ from the lack of $^{59}$Ni in CRs, 
as previosly reported by CRIS \cite{ACECRISOLD2}.
Combining this  two limits on $T$
implies that CRs are not accelerated by the same SN in which they are synthesized, 
and then CR acceleration must take place in  regions, like OB associations,  where at least two nearby supernovae occur within a few Myr.
\section{Electrons}
\label{s:electron}
\begin{figure}
\begin{center}
\subfigure[]
{
\includegraphics[height=6.5cm, width=5.4cm,angle=270]{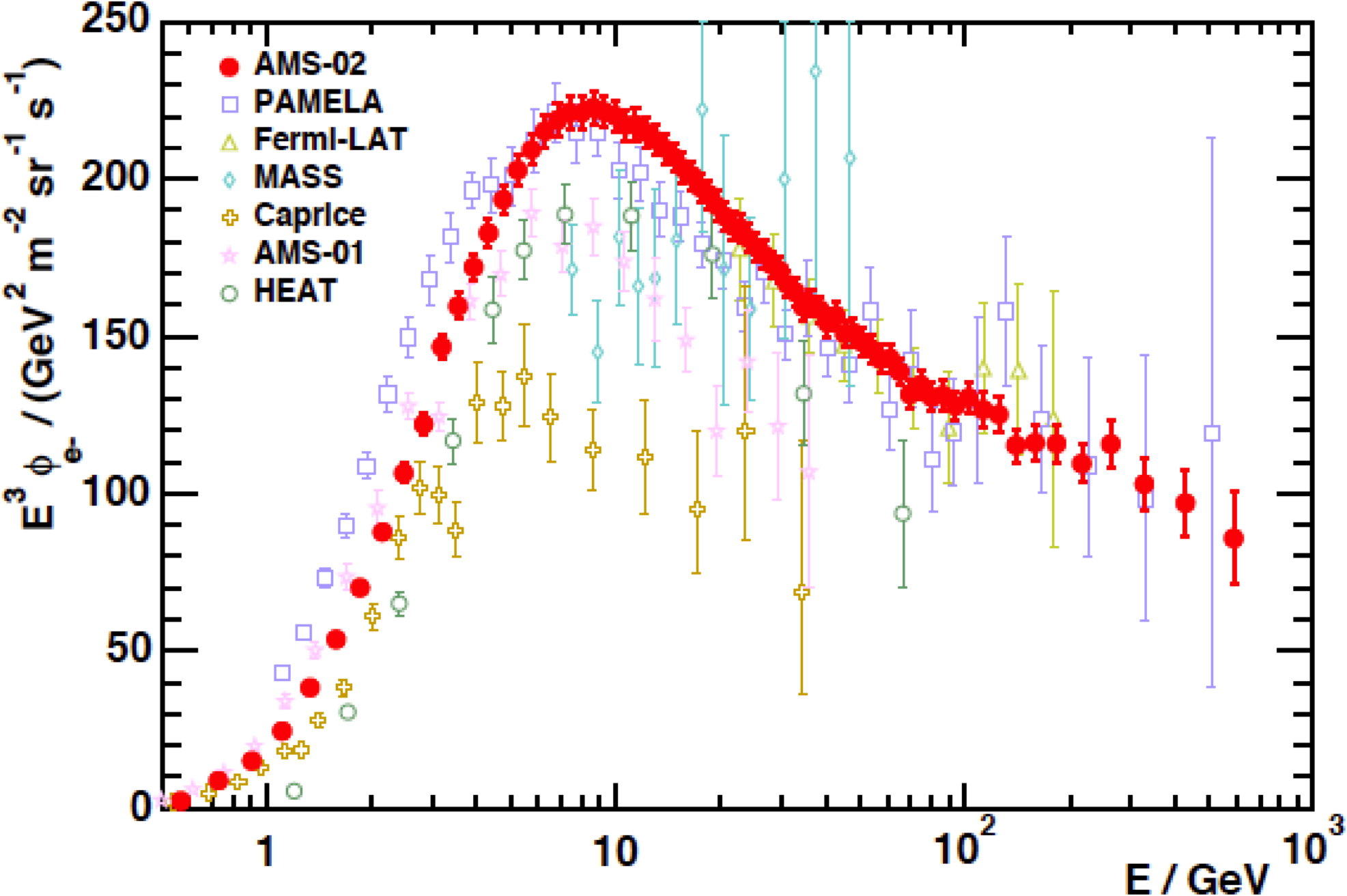}
\label{electron}                            
}
\hspace{10mm}
\subfigure[]
{
\includegraphics[height=6.5cm, width=5.4cm,angle=270]{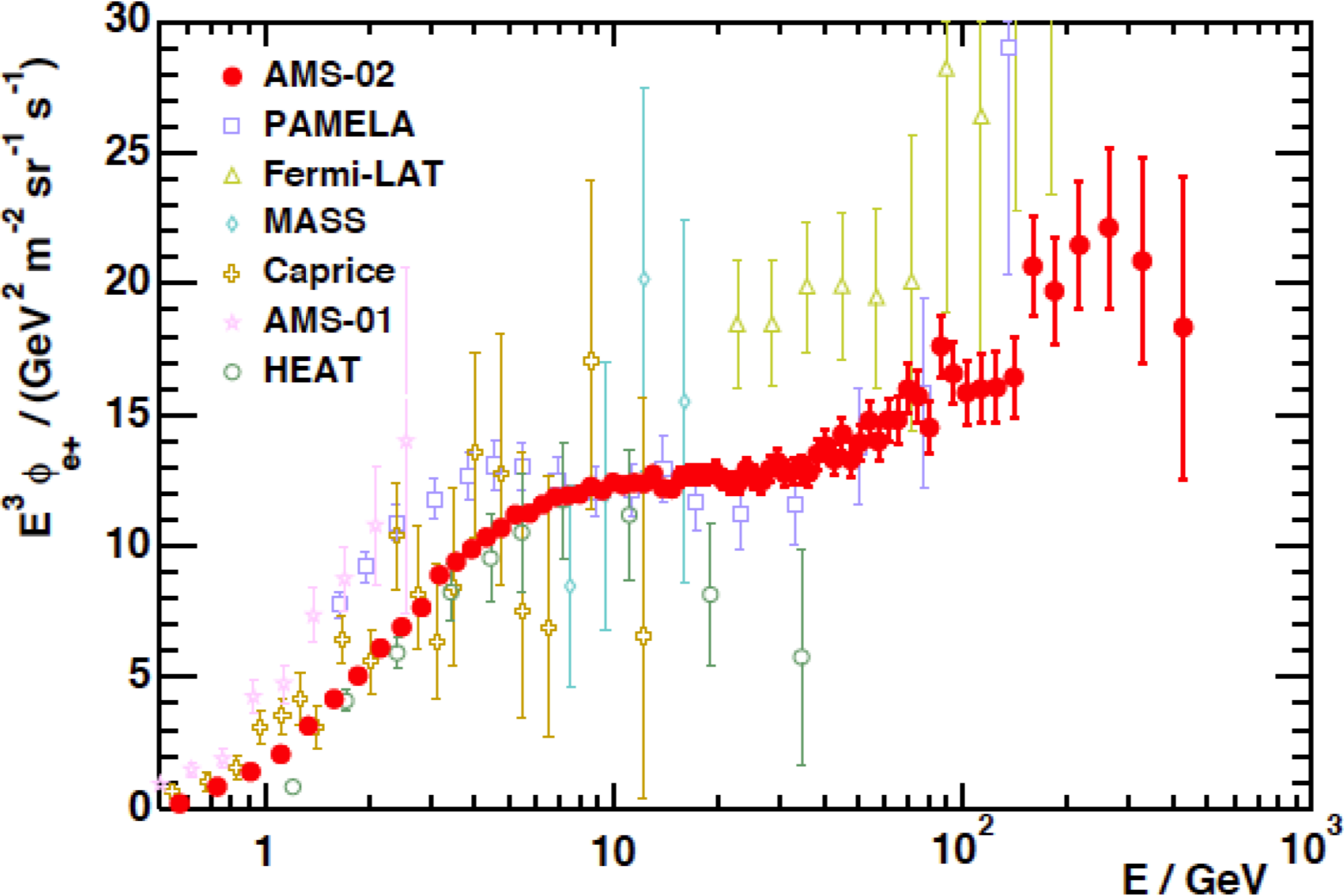}
\label{positron}                            
}
\vspace{5mm}
\subfigure[]
{
\includegraphics[height=6.5cm, width=5.4cm,angle=270]{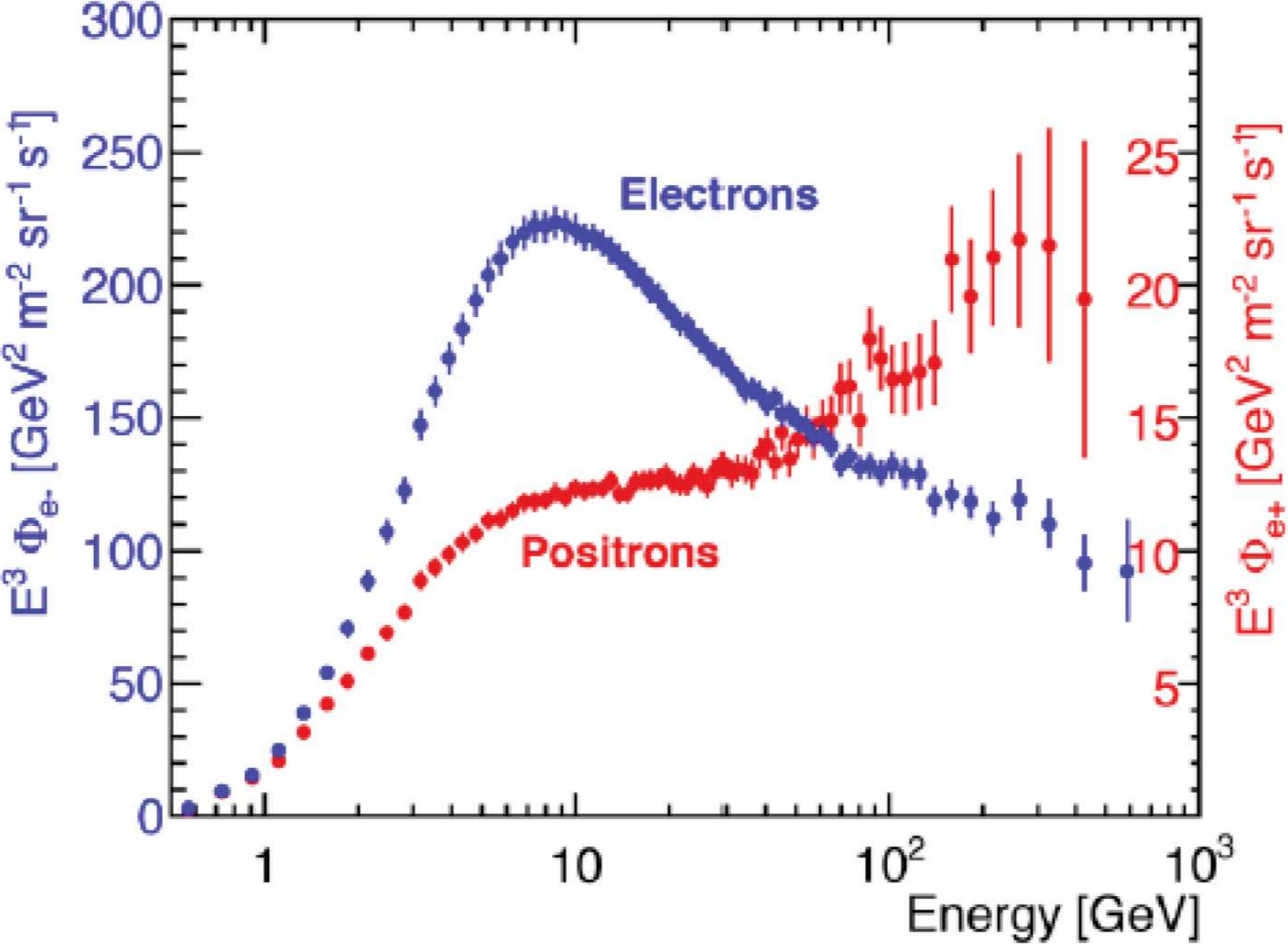}
\label{elepos_comp}                            
}
\hspace{10mm}
\subfigure[]
{
\includegraphics[height=6.2cm, width=5.0cm,angle=270]{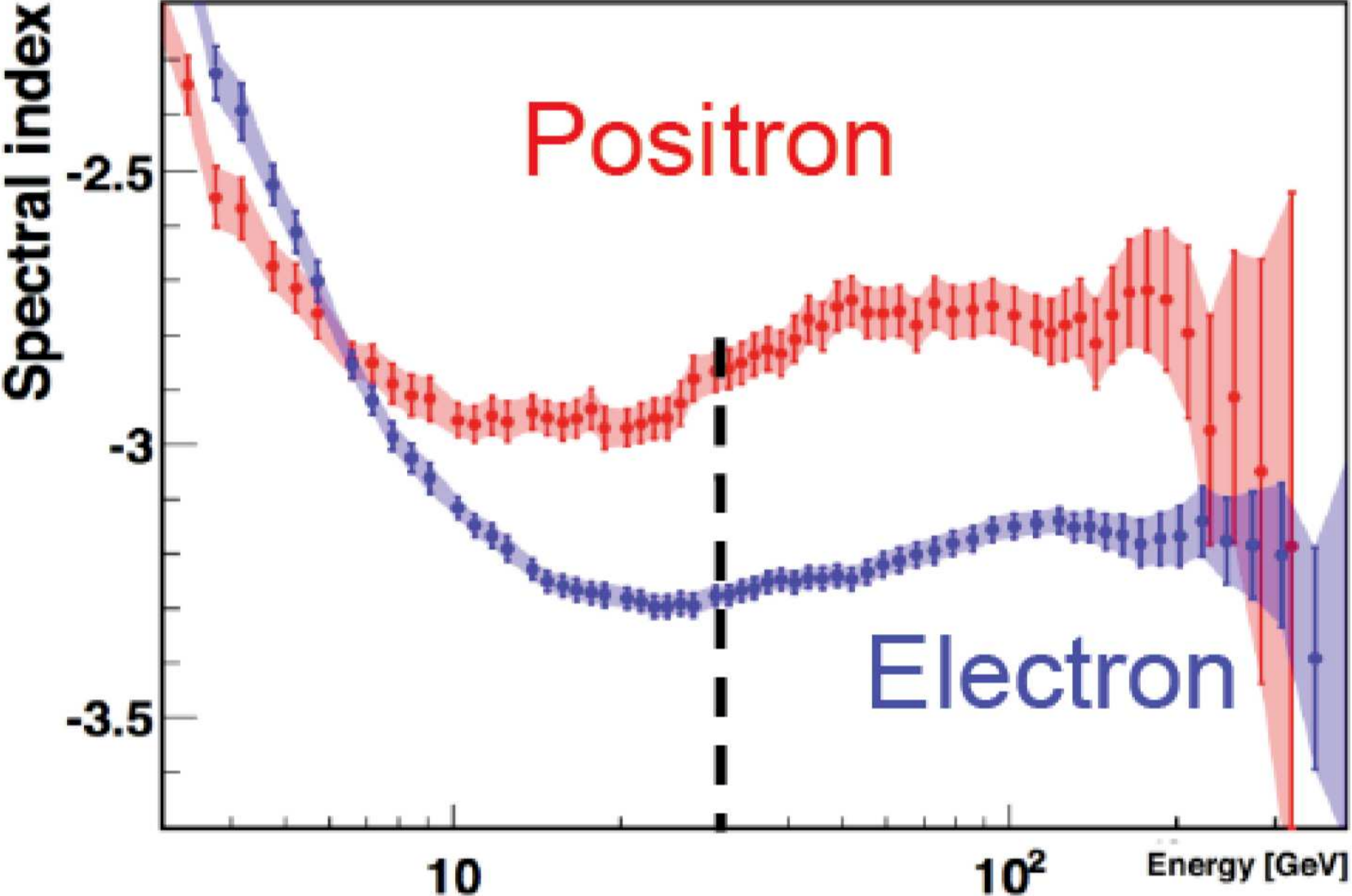}
\label{elepos_spectralindex}                            
}\vspace{-0.3cm}
\caption{(a) $e^-$ and (b) $e^+$ flux multiplied by $E^3$ as measured by AMS, 
compared with earlier measurements.
(c) Comparison of the $e^\pm$ fluxes measured by AMS. (d) The spectral indices $\gamma_{e^\pm}$ of the AMS $e^\pm$ fluxes  
 as a function of energy \cite{AMS-electronICRC,AMS-elepos}.
}
\label{figepm}  
\vspace{-0.4cm}
\end{center}
\end{figure}

The overall electron component ($e^++e^-$) accounts for about 1\% of CRs.
Secondary electrons and positrons are produced  in interactions 
of CR nuclei with the ISM, through the $\pi^\pm \rightarrow \mu^\pm \rightarrow e^\pm$ decay chains. 
Since  approximately equal amounts of $e^-$ and $e^+$ are expected from such processes, 
the observed overabundance of $e^-$ over $e^+$  
in CRs 
indicates that most of electrons have a primary origin.
Observations of X-ray and TeV gamma-ray emission from SNRs
provide clear evidence for the acceleration of electrons in SNR shock up to energies of about 100 TeV \cite{SNRaccele}.
Unlike the hadronic CR component, electrons, because of their low mass, suffer significant 
energy losses (proportional to their squared energy) during propagation in the Galaxy through 
synchrotron radiation in the galactic magnetic field and inverse Compton scattering off CMB photons.
These large energy losses explain the steeper electron spectrum compared to that of protons,
and place an upper limit on the age ($\sim$10$^5$ yr) and distance (<1 kpc) of the astrophysical sources
of TeV electrons that can be observed at Earth.
Since the number of such
nearby SNRs is limited, 
the electron energy spectrum above 1 TeV could exhibit spectral features 
and, at very high energies, a measurable anisotropy in the electron arrival
directions is expected, due to the locality of the possible sources. 
Possible excess in the electron spectrum could also result from 
a primary source of positrons, which seems necessary to explain the rise in positron fraction with energy, 
recently observed by PAMELA \cite{PAMELA-posfrac} and AMS \cite{AMS-posfrac}. Infact such a source, 
either of astrophysical (e.g. pulsars)
or exotic (e.g. dark matter) origin, would contribute with an equal number of positrons and electrons. \\
Measurements of the CR $e^\pm$ fluxes have 
been pursued since 1970's by  balloon-borne and space-based experiments. 
Because of the low intensity of the $e^{\pm}$ signals
and the large CR proton background, the main requirements
for these instruments are 
a large acceptance, long exposure time and an excellent $e/p$ discrimination capability.
Calorimeters can be used to measure the ``all'' electron ($e^++e^-$) spectrum, 
while separation between $e^+$ and $e^-$  makes mandatory the
use of a magnetic spectrometer to determine the sign of the charge. \\
In recent years, a great interest has arisen from several new electron measurements.
ATIC-2 reported an excess in the all electron spectrum  at energies between 300 and 800 GeV \cite{ATIC-ele}, 
which was not confirmed by subsequent measurements by Fermi and HESS.
Fermi data follow a power-law spectrum with spectral index -3.08$\pm$0.05 \cite{FERMI}, consistent with HESS result, 
which moreover reported a rapid steepening above 1 TeV \cite{HESS}. 
\begin{figure}
\begin{center}
\subfigure[]
{
\includegraphics[height=6.6cm, width=5.4cm,angle=270]{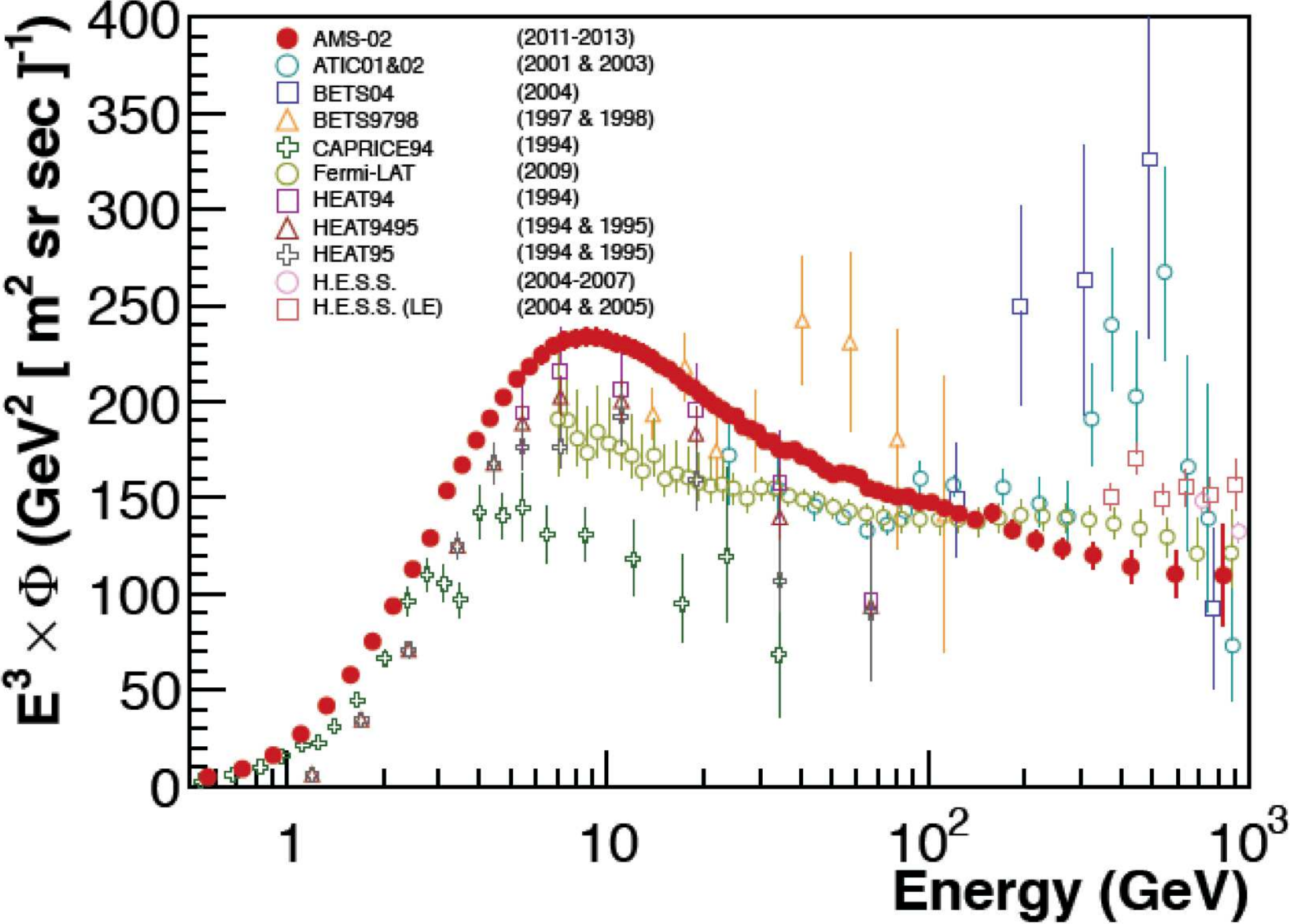}
\label{eleposComp}                            
}
\hspace{10mm}
\subfigure[]
{
\includegraphics[height=6.6cm, width=5.4cm,angle=270]{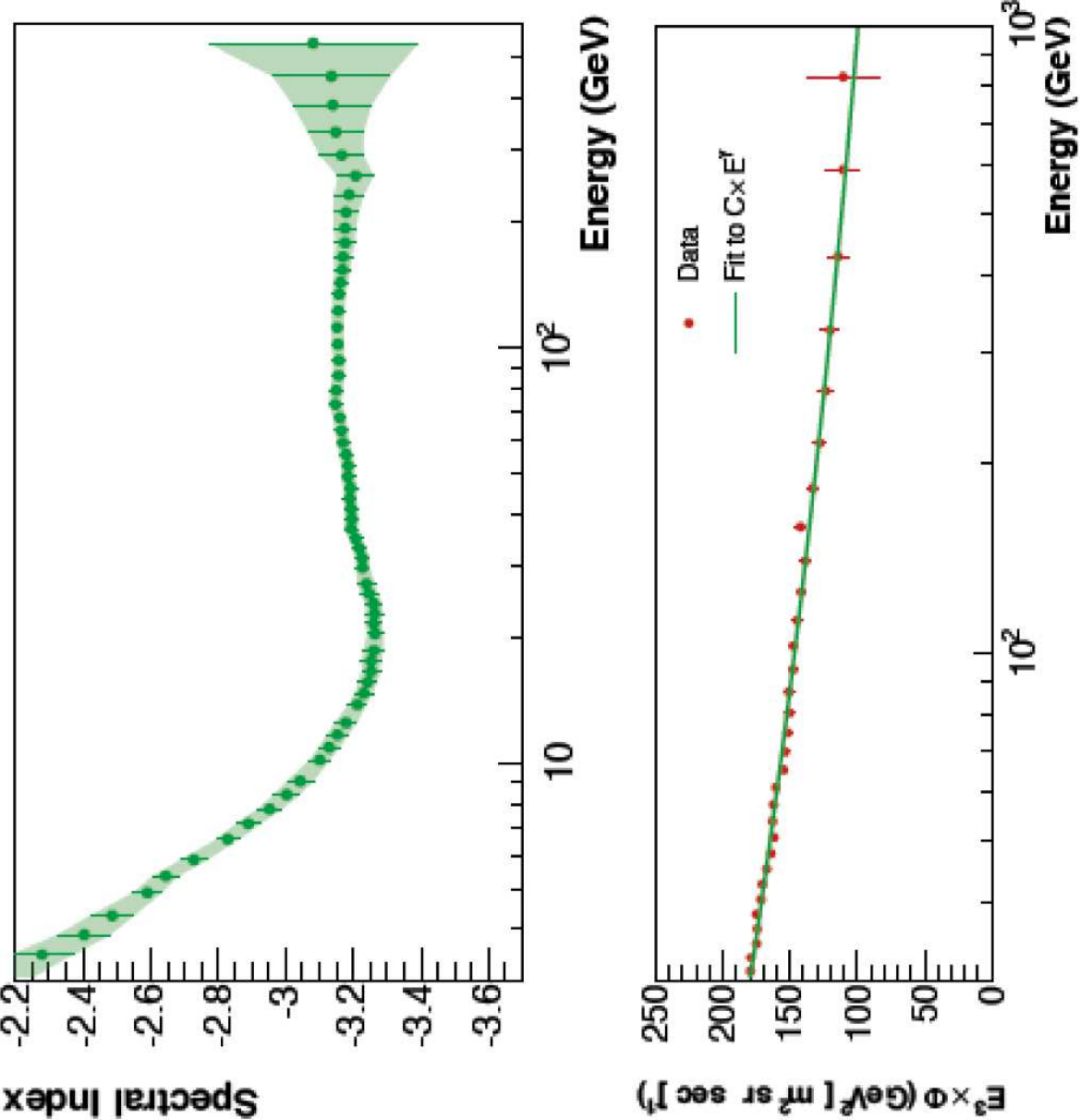}
\label{eleposfit}                            
}
\caption{(a) The AMS $e^++e^-$ flux 
multiplied by $E^3$ as a function of energy $E$, compared with earlier results.  
(b) Top: spectral index of the all electron flux as a
function of energy. Bottom panel:
single power-law fit to the AMS $e^++e^-$ flux multiplied by $E^3$ above 30.2 GeV \cite{AMS-allelectron}.
}
\label{elepos}
\vspace{-0.8cm}
\end{center}
\end{figure}
At this ICRC, AMS reported precision measurement of the fluxes of CR $e^-$, $e^+$ and  $e^++e^-$ \cite{AMS-electronICRC}.
Unlike the $p$ and He analyses, the energy of electrons  is measured with the ECAL, and the energy bins
are chosen according to its energy resolution ($\sim$2\% at 300 GeV). 
Statistical estimators, based on Boosted Decision Tree and log-likelihood algorithms, 
have been developed to discriminate $e^\pm$ from $p$ by exploiting their different
shower shapes in ECAL and energy deposited in TRD proportional tubes. 
The $p$ rejection power of ECAL is $\sim$10$^4$ above few GeV for  90\% $e^\pm$  selection efficiency, 
while for TRD it reaches 10$^4$ at 30 GeV and decreases rapidly above.
In each energy bin, signal and background yields are obtained with a data-driven template-fit approach.
The $e^++e^-$ flux is not computed as the sum of $e^-$ and $e^+$ fluxes,
but with a dedicated analysis 
that allowed to reduce the systematic uncertainties 
and increase the statistics of the sample at higher energies.
In 30 months of operations, AMS identified
10.6$\times$10$^6$ $e^++e^-$ events  between 0.5 GeV and 1 TeV.
The resulting flux (Fig.~\ref{eleposComp}) is consistent with a single power-law above 30 GeV with spectral index 
$\gamma$ =-3.170$\pm$0.008 (stat+syst)$\pm$0.008 (energy scale), 
as shown in Fig.~\ref{eleposfit}.
No structures are observed in the $e^++e^-$ spectrum \cite{AMS-allelectron}.\\
As far as the $e^-$ spectrum is concerned, only in recent years 
CR $e^-$ above 50 GeV were identified for the first time by PAMELA, which
measured the $e^-$ spectrum between 1 and 625 GeV. 
This is well described by a single power-law energy dependence with spectral index -3.18$\pm$0.05 
above the energy of 30 GeV, and shows
no significant spectral features within the errors \cite{PAMELA-ele}. \\
A precise measurement of the $e^-$ flux was presented by AMS (Fig.~\ref{electron}),
based on the identification of 9.23$\times$10$^6$ $e^-$ events in the energy range 0.5-700 GeV \cite{AMS-electronICRC}.
The AMS $e^-$ spectrum cannot be fitted to a single power-law over the entire energy range
not affected by solar modulation (>10 GeV).
It turned out that the spectral index $\gamma_{e^-}$ has an energy dependence (Fig.~\ref{elepos_spectralindex})
and hardens at $\sim$30 GeV \cite{AMS-elepos}.
Further discussion of this point invokes the comparison with the $e^+$ flux behaviour and is 
presented in section \ref{s:positrons}.\\
The multi-TeV region, where there is high potential for studying local cosmic accelerators, 
is still largely unexplored.
Thanks to their large collection area, 
Imaging Atmospheric Cherenkov Telescopes (HESS, MAGIC) proved to be able  to 
extend the  $e^++e^-$ spectrum beyond 1 TeV with large statistics, 
at the expense of a quite uncertain identification of the CR particle inducing the air shower.
In \cite{VERITAS},
 the ground-based experiment VERITAS reported a measurement of the electron spectrum between 300 GeV and 4 TeV.
The electron signal extraction is strongly based on Monte Carlo simulations
and multivariate analysis methods, resulting in 
large systematic uncertainties on the energy scale, and significant contamination from 
gamma-ray events and protons. 
VERITAS data confirms a cutoff in the spectrum around 1 TeV, first observed by HESS \cite{HESS}, 
above which the spectral index rapidly changes from $\approx$-3.2  to $\approx$-4.1.\\
An interesting attempt to measure directly the electron flux at several TeV 
exploiting a novel detection technique was presented by the balloon-borne experiment 
CREST (Cosmic Ray Electron Synchrotron Telescope) \cite{CREST}.
Synchrotron X-rays emitted from TeV electrons bending in the geomagnetic field
are detected by means of an array of BaF$_2$ crystals readout by photomultipliers
and surrounded by  a plastic scintillator veto system. 
Electrons are identified by the characteristic signature of synchrotron photons, which form 
a co-linear, isochronous hit pattern in the detector. 
The effective acceptance  of the instrument is very large (18.5 m$^2$sr at 5 TeV), 
because only photons, and not directly electrons, are required to cross the detector.
Moreover it increases with the electron energy, as the synchrotron photon yield. 
CREST was flown for 10 days over Antarctica in 2011. Unfortunately, 
the measurement was affected by 
the high level of photon background due to secondary production in the atmospheric
overburden, which overwhelmed the expected synchrotron signals. 
Only a preliminary  upper limit on the integral electron flux 
(<7.11$\times$10$^{-3}$m$^{-2}$sr$^{-1}$s$^{-1}$) was calculated at energies >15 TeV. \\
Finally, on the low-energy side, 
it is worth recalling 
 the Voyager 1 measurement of the unmodulated electron spectrum  between 10 and 100 MeV in the local ISM 
\cite{Cummings}.
\section{Antimatter}
\label{s:antimatter}
Positrons and antiprotons are rare components of CRs.
Since they are thought to be mainly of secondary origin, i.e. they are generated through collisions of CR nuclei with atoms of the ISM, 
detailed measurements of their energy spectra 
can provide important information about the CR propagation mechanisms.  
However the recent data from PAMELA and AMS
show unexpected features in these spectra that require new interpretations.
\subsection{Positrons}
\label{s:positrons}
\begin{figure}
\begin{center}
\subfigure[]
{
\includegraphics[height=6.6cm, width=5.4cm,angle=270]{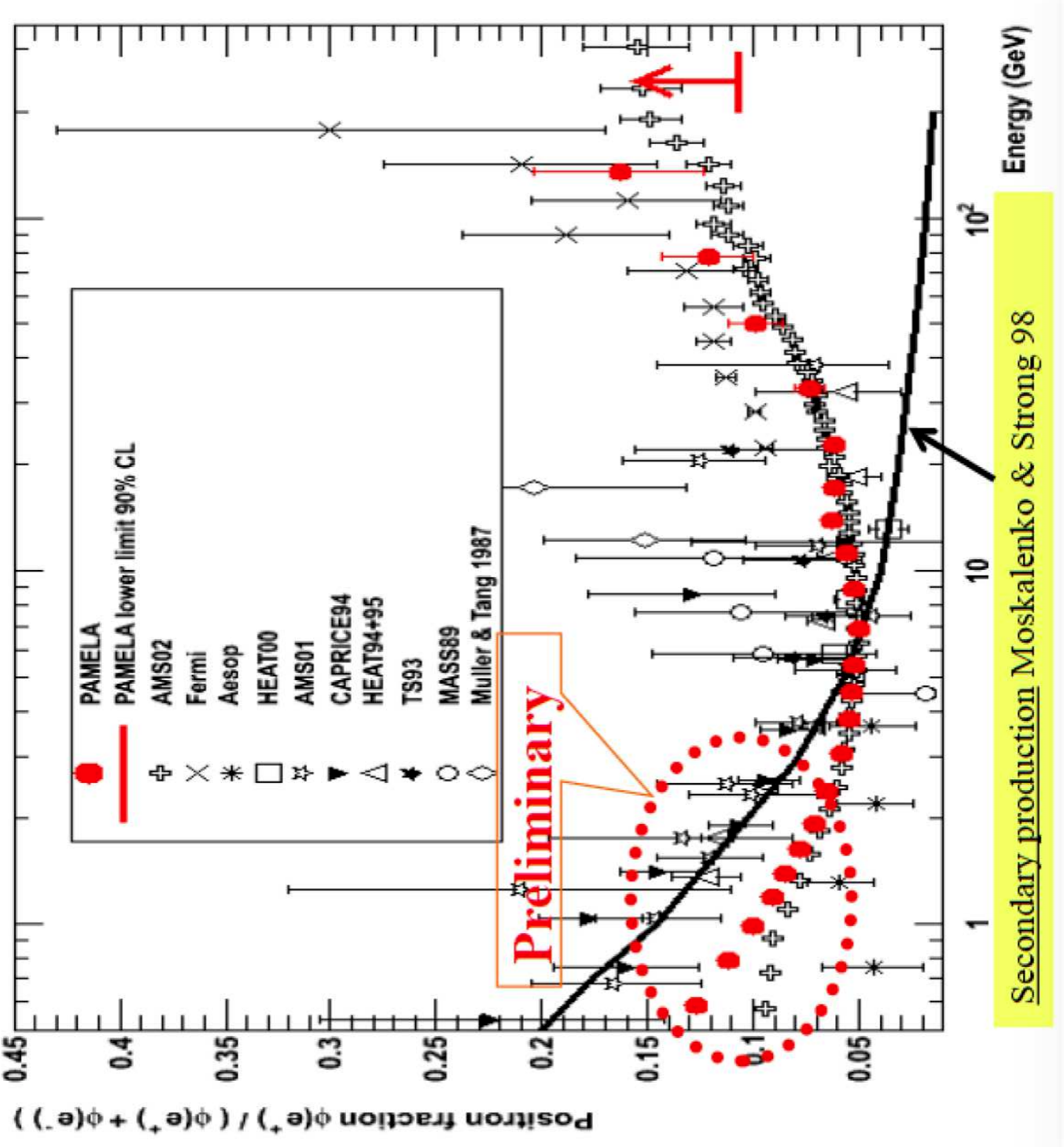}
\label{posfrac_a}                            
}
\hspace{8mm}
\subfigure[]
{
\includegraphics[height=6.8cm, width=5.4cm,angle=270]{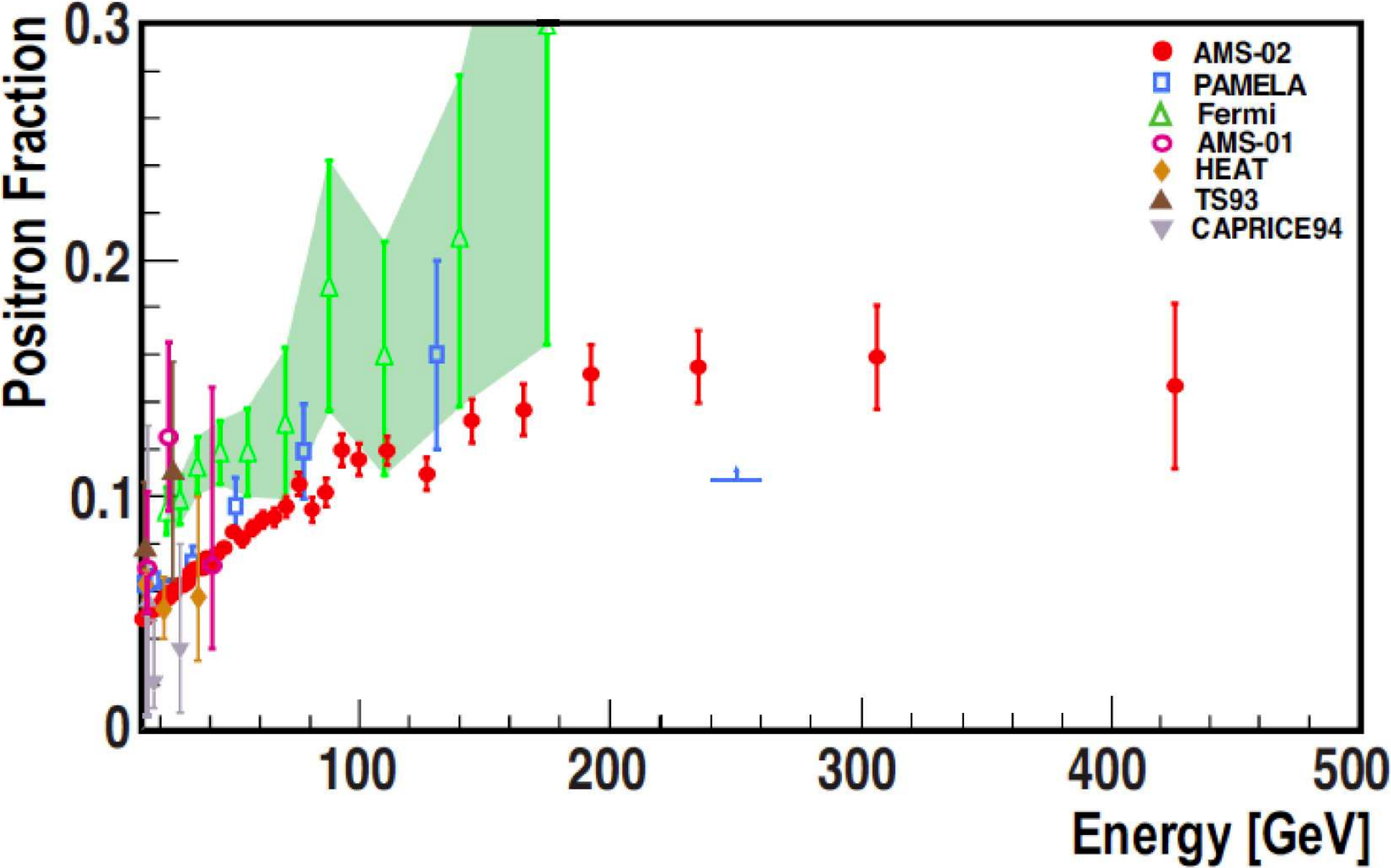}
\label{posfrac_b}                            
}
\caption{(a) The positron fraction in the energy range 0.5-200 GeV. Preliminary unpublished 
measurements by PAMELA in the low energy region (<2 GeV) were presented at the ICRC \cite{PAMELAhl}. 
The low-energy discrepancy between PAMELA and AMS with data collected during the 1990's
is interpreted in terms of charge-sign solar modulation effects \cite{PAMELA-pos}.
(b) The AMS positron fraction as a function of energy above 10 GeV, where it begins to increase, 
compared with PAMELA and Fermi results \cite{AMS-posantipICRC}.
}
\label{posfrac}
\vspace{-0.8cm}
\end{center}
\end{figure}

A major experimental challenge 
in the $e^+$ measurement is represented by the need to 
reject the overwhelming 
background of protons ($p/e^{+}$ flux ratio $\sim$10$^3$ at 10 GeV) and electrons ($e^{-}/e^{+}\sim$10 at 10 GeV).
Discrepant results were reported by pioneering experiments, which were 
affected by a significant misidentification of the proton background.
In the 1990's, the balloon-borne magnetic spectrometers HEAT\cite{HEAT} and CAPRICE \cite{CAPRICE}, 
and AMS-01 in space \cite{AMS01} 
 measured the positron fraction $e^+/(e^++e^-)$ up to 30 GeV, though with limited statistics above 10 GeV. 
A predominantly decreasing positron
fraction with increasing energy was observed with a small excess around 7 GeV.
The first measurement of  $e^+/(e^++e^-)$ reaching 100 GeV  with
significantly better precision than earlier data, was reported by PAMELA \cite{PAMELA-posfrac}.
It clearly shows a rise of the positron fraction above 10 GeV (Fig.~\ref{posfrac_a}), 
in conflict with secondary production mechanisms, predicting a decresing trend with energy. 
Fermi LAT, though not equipped with a magnet, confirmed the positron rise between 20 and 200 GeV, with large
systematic uncertainties, 
by exploiting the Earth's shadow to separate $e^-$ from $e^+$ \cite{FERMI2}. \\ 
In these ICRC papers \cite{AMS02, AMS-electronICRC, AMS-posantipICRC}, 
AMS presented its recently published results 
of the positron fraction  and $e^+$ flux. The ratio
$e^+/(e^+ + e^-)$ was measured with unprecedented precision from 0.5 to 500 GeV\cite{AMS-posfrac}.
It steadily increases from 10 to $\sim$250 GeV then reaching a maximum at 275$\pm$32 GeV (Fig.~\ref{posfrac_b}). 
At higher energies, the fraction is no longer increasing but 
the data are still statistically limited to understand its behaviour.
The observed rise in the positron fraction is due to an excess of positrons 
as pointed out by comparing the individual AMS $e^-$ and $e^+$ fluxes, shown in Fig.~\ref{figepm}.
Similarly to $e^-$ flux, the $e^+$ flux cannot be described by a single power-law over the entire energy range;
its spectral index $\gamma_{e^+}$ shows an energy dependence and hardens at $\sim$30 GeV. 
Moreover the spectral indices for $e^+$ and $e^-$
are significantly different with
$\gamma_{e^-}$ < $\gamma_{e^+}$ between 20 GeV and 200 GeV (Fig.~\ref{elepos_spectralindex}), 
i.e. the electron flux is softer than the  positron one.
This demonstrates that the increase with energy observed in the
positron fraction is due to the hardening of the positron
spectrum and not to the softening of the electron spectrum above 10 GeV  \cite{AMS-elepos}.\\
The sharp rise observed in the positron fraction prompted
many theoretical works trying to explain the excess of $e^+$, 
either including additional sources of high-energy $e^\pm$ 
or reassessing critically the secondary $e^\pm$ production from CR interactions with ISM. 
As far as this latter case is concerned, it was shown, for instance, that by tuning the CR propagation parameters \cite{Blum}, 
or with inhomogeneous diffusion models \cite{Tomassetti1}, 
is still possible to reproduce the data under the assumption of purely secondary production.
In \cite{Blasi2}, the positron excess is explained 
taking into account  $e^\pm$  produced as 
secondaries in the SNRs where CRs are accelerated.
Conversely, astrophysical objects like nearby pulsars,
or dark matter (DM) particles have been proposed as possible primary sources of positrons. 
In \cite{ICRCBoudad}, both these hypotheses have been tested against the recent AMS data. 
They showed that 
leptophilic DM is disfavoured, while 
DM annihilation in quarks and gauge or Higgs bosons would be able to produce the positron excess, 
but only assuming high cross-sections 
not compatible  with constraints from other observations.
Instead an explanation based on a single pulsar seems more feasible,  
and five known nearby objects are identified which can fit the data.
In \cite{ICRCDiMauro}, the interplay between astrophysical sources and DM is studied, 
showing that most of the primary $e^+$ component can be produced  by
Pulsar Wind Nebulae (PWNe), with possible sub-leading components from SNRs and DM (Fig.~\ref{posfrac_dimauro}). 
In \cite{ICRCRozza}, a better fit to the AMS $e^++e^-$ spectrum above 100 GeV is obtained 
by adding a contribution from the pulsars Vela-X and Monogem to the standard 
local interstellar spectrum calculated with GALPROP.\\
However, the uncertainties on CR propagation parameters significantly affect 
the predicted positron rise for both the scenarios \cite{ICRCBoudad2}, thus
strongly limiting at the moment the possibility to ascertain the origin of the excess.
It was remarked that 
a different shape of the positron fraction above 500 GeV 
is expected depending on the astrophysical or DM origin of the $e^+$ excess.
That might be explored with future AMS data.
Moreover in case of a  few pulsar sources, a dipole anisotropy of order 1\%  should be observed
at hundreds of GeV.  
So far, AMS measured a $e^+/e^-$ ratio consistent with isotropy \cite{AMS-posfrac}, while 
anisotropies searches separately for  $e^\pm$ are underway \cite{AMS-anisotropy1}.
A study of the arrival directions of $e^+$ and $e^-$ taking into account the effects of the Earth's geomagnetic field
was presented by PAMELA. Results are consistent with isotropy at all angular scales considered.
An upper limit of 0.076 at 95\% confidence level on the amplitude
of dipole anisotropy  is calculated for positrons \cite{PAMELA-anisotropy}. \\
Several works based on analyses of PAMELA data at low energies (<10 GeV)
have been also presented at the ICRC, as for instance
the  study of the time variation of $e^\pm$  fluxes  over the 23$^{rd}$ solar minimum  \cite{PAMELA-solarele}
and of the positron fraction betwen 2006 and 2015, aimed at investigating  charge-sign 
dependent solar modulation effects;  and 
the study of the secondary $e^\pm$ 
stably trapped and albedo components produced in interactions of CR protons with the
residual atmosphere \cite{PAMELA-solarfrac}.
\subsection{Antiprotons}
\label{s:antiproton}
The major difficulty in measuring the antiproton
flux is the positive identification of the rare antimatter 
particles in the presence of a large background,
mostly consisting  of electrons and, for experiments performed with balloons at 
small atmospheric depths, secondary 
pions and muons, which are both
produced in the interaction with the atmosphere.
%
\begin{figure}[]
\begin{center}
\subfigure[]
{
\includegraphics[height=6.6cm, width=5.4cm,angle=270]{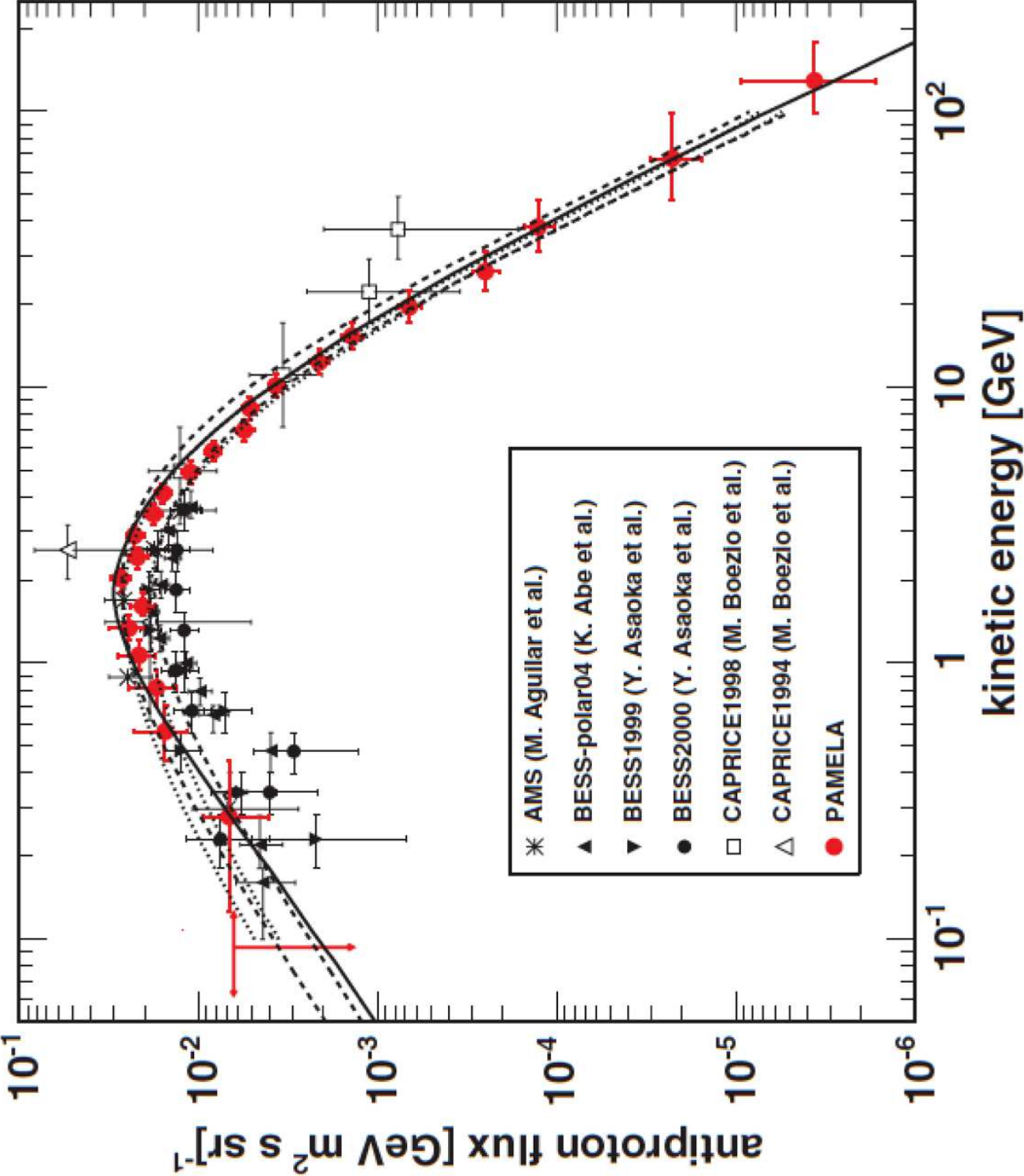}
\label{antip_a}                            
}
\hspace{8mm}
\subfigure[]
{
\includegraphics[height=6.8cm, width=5.4cm,angle=270]{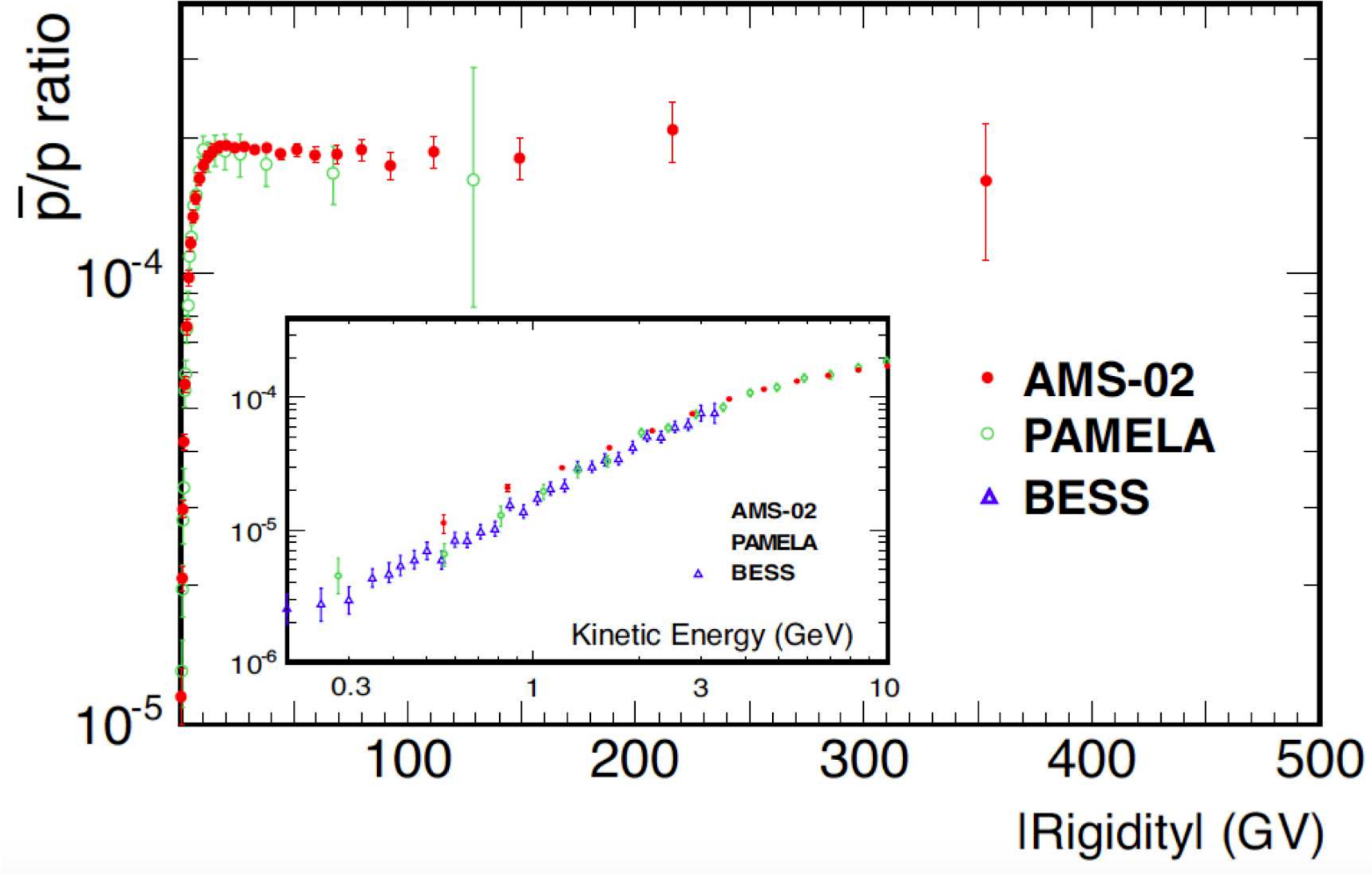}
\label{antip_b}                            
}
\caption{(a)  $\bar{p}$  flux measured by PAMELA, compared with previous results and 
theoretical calculations for a pure secondary $\bar{p}$ production \cite{PAMELA-antip}.
(b) AMS $\bar{p}/p$  ratio up to 450 GV \cite{AMS-posantipICRC}, 
compared with PAMELA \cite{PAMELA-antip} and BESS-Polar \cite{BESSPOLARII-antip} results. 
An expanded view of the low-energy region is shown in the inset plot.
}
\label{antip}
\vspace{-0.5cm}
\end{center}
\end{figure}
Early measurements of antiprotons were severely  compromised by large background contamination 
due to an inadequate particle identification. 
Magnetic spectrometers in 1990's measured the antiproton flux below 10 GeV
and only CAPRICE98 \cite{CAPRICE98_pbar}  reached few tens of GeV. 
BESS95+97  showed that the  $\bar{p}$ spectrum has
a distinct peak at 2 GeV  with the flux decreasing  at lower energies, as expected 
from the kinematic constraints of $\bar{p}$ production in the ISM \cite{BESS_9597}.
That confirmed that the secondary component is dominant in the CR $\bar{p}$ flux.
However a possible  excess at energies lower than the secondary peak, which was not ruled out by BESS95+97 data, 
could result  
from cosmologically primary sources of $\bar{p}$, like
evaporation of primordial black holes (PBH).
This exotic component was later excluded by BESS-Polar II precise measurement between 0.17 and 3.56 GeV
near solar minimum in Dec. 2007 \cite{BESSPOLARII-antip}. 
The first observations of both $\bar{p}$ flux (Fig.~\ref{antip_a}) and 
$\bar{p}/p$ flux ratio  extending to high energy (180 GeV) were performed by PAMELA \cite{PAMELA-antip}.
At the ICRC, AMS presented  a preliminary measurement 
of the $\bar{p}/p$ ratio, based on the identification of 2.9$\times$10$^5$ $\bar{p}$ selected in the rigidity
range 1-450 GV \cite{AMS-posantipICRC}. This result has unprecedented precision and clearly shows that
the $\bar{p}/p$ ratio is almost constant with rigidity above 100 GV (Fig.~\ref{antip_b}).\\
This unexpected behaviour at high energy has triggered speculations 
about a possible primary antiproton component originating from annihilation of DM particles.
At the beginning, a misleading comparison of AMS data with old secondary production predictions (e.g. \cite{Donato}), 
which indicated a rapid decrease of the $\bar{p}/p$ ratio with increasing energy, 
seemed to point to the DM scenario.
\begin{figure}[b]
\begin{center}
\subfigure[]
{
\includegraphics[height=6.6cm, width=5.4cm,angle=270]{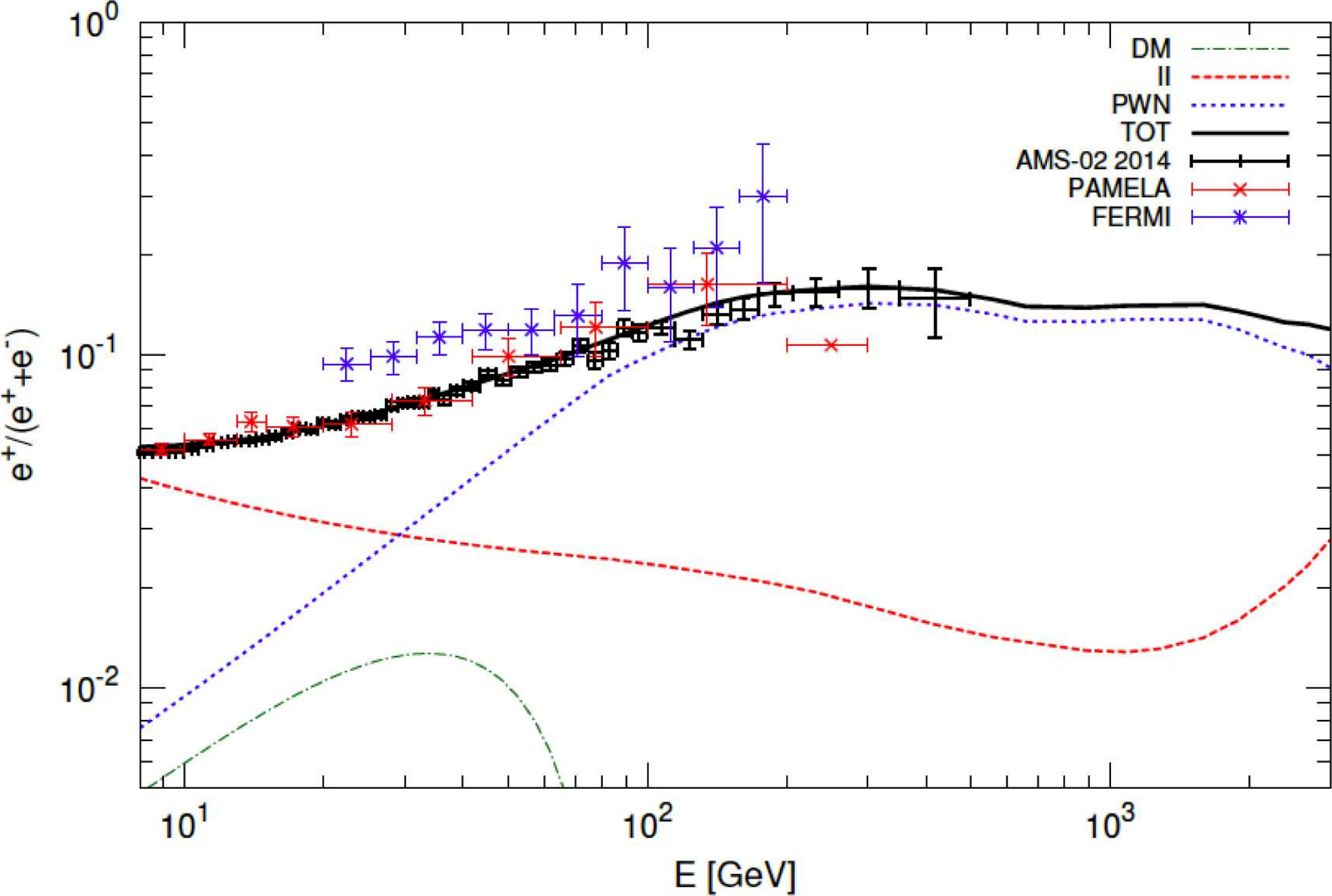}
\label{posfrac_dimauro}                            
}
\hspace{7mm}
\subfigure[]
{
\includegraphics[height=6.6cm, width=5.4cm,angle=270]{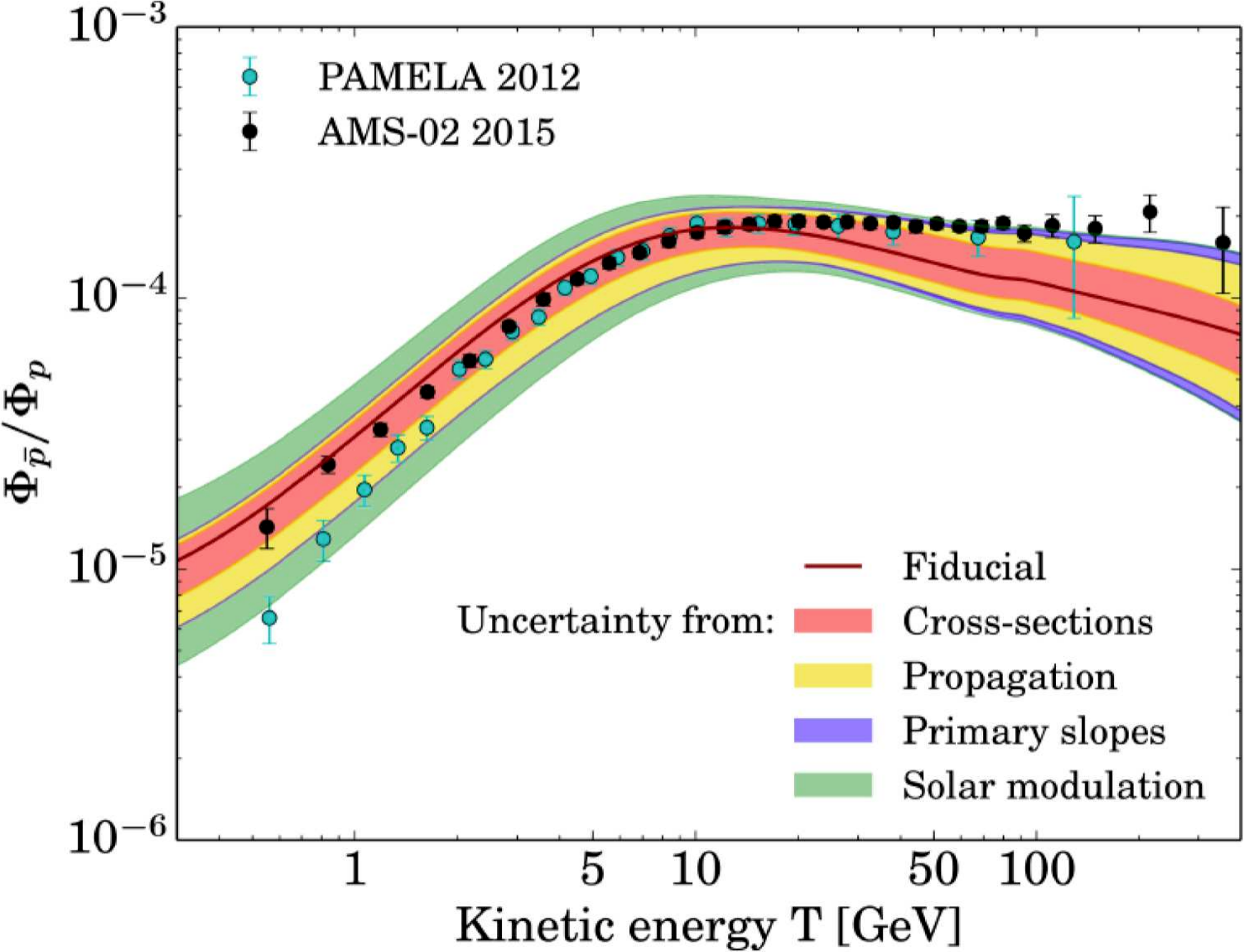}
\label{antip_boudad}                            
}
\caption{(a) Prediction for the positron fraction calculated in \cite{ICRCDiMauro}, 
including contributions from  secondaries, PWNe and DM annihilating into the $\mu^+\mu^-$ channel. (b) The combined total uncertainty on the secondary $\bar{p}/p$ ratio calculated in \cite{ICRCGiesen}, superimposed to PAMELA and AMS data.
}
\label{theory}
\end{center}
\end{figure}
Hence in order to clarify the origin of the anomaly, 
calculations of the secondary $\bar{p}/p$ flux ratio have been reassessed
using updated propagation models, taking into account
additional effects previously neglected (e.g. $\bar{p}$ energy losses (including tertiary component), 
diffusive reacceleration, solar modulation), 
harder $p$ and He spectra at high energy, 
 diffusion coefficient consistent with recent B/C data, 
newer data on $\bar{p}$ production cross sections.
Several theoretical works have been presented driven by this approach 
and aimed at a critical evaluation of the relevant uncertainties affecting the computations 
\cite{Donato2,ICRCGiesen,ICRCCowsik,Kappl,Evoli}. An example in shown in Fig.~\ref{antip_boudad}.
They all come to a  similar conclusion: that 
the  flat $\bar{p}/p$ ratio at high energy is consistent with pure secondary 
production within the error band of the updated predictions. 
However, given the current uncertainties, mainly on propagation parameters, 
a possible $\bar{p}$ contribution from DM cannot be ruled out.  
More accurate $\bar{p}$ data at high energy, expected from AMS in next years, 
combined with an improved modeling of the secondary production, 
will allow to search with better sensitivity for possible sub-dominant DM signals 
in a large background of secondary $\bar{p}$.
\section{Future experiments}
\label{s:futurexp}
Several new experiments have been presented at the ICRC.
Some of them are already  operating \footnote{at the time of writing} (NUCLEON, CALET) or 
foreseen to start taking data soon (DAMPE, ISS-CREAM, CSES), 
others have only been proposed (HNX, HELIX, GAPS) or 
they are in an initial phase of design and development (GAMMA-400).
Some experiments address  broad observation targets (CALET, NUCLEON, DAMPE, ISS-CREAM), 
while others are more focused on specific measurements (CSES, HNX, HELIX, GAPS). 
It can be noticed that 
the projects that are expected to produce results in the next few years (CALET, NUCLEON, ISS-CREAM and DAMPE)
are all based on calorimetric instruments, featuring 
large depth and acceptance and long exposure aimed at  
measuring the individual spectra of CR nuclei up to hundreds of TeV approaching the knee, and the all electron spectrum 
up to several TeV. 
Then they are foreseen to  provide overlaps with AMS $p$, He, and $e^++e^-$ spectra  
and  extend such measurements well beyond the energy limit imposed by the MDR of the spectrometer.
In the following paragraphs, I will describe the main experiments 
which will likely present new results at the next ICRC, comparing  their 
detection techniques and performance, observation targets and expected outcome. 
Finally, the most interesting proposals presented at the ICRC will be briefly sketched. 
\subsection{NUCLEON}
NUCLEON is a russian spaced-based detector  aimed at measuring the
energy spectra of individual CR nuclei from $Z$=1 to $Z$=30 in the energy range from 100 GeV up to 1 PeV. 
It was successfully launched on a
RESURS-P satellite on Dec. 26$^{th}$ 2014 and it is planned to take data for 5 years.
The instrument includes 
four silicon pad  detectors to measure the particles charge,  a carbon 
target, six scintillator layers for the trigger system, 
a tracker 
made of six layers of Si microstrips interleaved with thin W layers,   
and a Si microstrip-Tungsten sampling  calorimeter. 
The total thickness is about 16 radiation lengths ($X_0$).
The energy of CR particles is determined by two independent techniques: 
a Kinematic Lightweight Energy Method (KLEM), 
based on the measurements of the spatial density of secondary shower particles produced in the carbon target and W layers 
by means of the Si microstrip tracker, and a traditional ionization calorimeter technique. 
The sampling calorimeter can measure also electrons in the energy range 100 GeV-3 TeV.
The effective geometric factor is more than 0.2 m$^2$sr for nuclei and 0.06
m$^2$sr for electrons \cite{NUCLEON}.
\subsection{CALET (CALorimetric Electron Telescope)}
CALET is a space mission led by the Japanese Space Agency (JAXA)
with the participation of the Italian Space Agency (ASI) and NASA. 
CALET was launched on August 19$^{th}$, 2015  by the Japanese rocket  H-II Transfer Vehicle (HTV-5)
and robotically installed on the Japanese Experiment Module Exposure Facility (JEM-EF) of the ISS
on August 24$^{th}$, for a two-year mission, 
with a possible first extension to 5 years \cite{CALET}.
Its main scientific goal is to search for possible clues of the
presence of astrophysical sources of high-energy electrons nearby the Earth or
signatures of DM, by measuring accurately the $e^-+e^+$ spectrum from 1 GeV up
to several TeV \cite{CALET6}. CALET will also
measure the energy spectra and elemental composition of CR nuclei from H to Fe
up to hundreds of TeV, and the abundance of UHGCRs  at
few GeV/amu up to about Z=40  \cite{CALET5}. 
The instrument consists of two layers of
segmented plastic scintillators to identify the particle charge, a thin tungsten-scintillating fiber 
imaging calorimeter providing accurate particle tracking and identification by multiple dE/dx sampling, and a thick PWO
crystal calorimeter to measure the energy of CRs with excellent resolution and
electron/hadron separation up to the multi-TeV scale. 
The total thickness is equivalent to 30 $X_0$ and 1.3 proton interaction lengths ($\lambda_I$). 
The geometrical factor is 0.12 m$^2$sr and the total weight is 613 kg. 
An extensive campaign of beam tests for calibration was carried out at GSI and CERN-SPS from 2011 to 2015
with beams of accelerated electrons, muons, protons and ion fragments \cite{CALET2}. 
The main  features of the detector are:
a charge resolution ranging from 0.15  for B to 0.30-0.35 charge units in the Fe region;
an energy resolution better than 2\% for electrons above 100 GeV \cite{CALET3};
an angular resolution of 0.1$^\circ$ for electrons and better than 0.5$^\circ$ for hadrons \cite{CALET8};
a good energy linearity for electrons and hadrons up to several TeV \cite{CALET4};
a proton rejection power $\sim$10$^5$ with 80\% electron efficiency \cite{CALET7}.
\subsection{ISS-CREAM (Cosmic Ray Energetics And Mass)}
ISS-CREAM have been built transforming the  CREAM payload,
successfully flown in six flights 
over Antarctica, for accommodation on the ISS JEM-EF for a 3 year mission  \cite{ISSCREAM}. 
The exposure will be increased by an order of magnitude allowing ISS-CREAM to extend
the measurement of the energy spectra of CR nuclei 
 up to hundreds of TeV.
The ISS-CREAM payload ($\sim$1300 kg) is currently being integrated at NASA Wallops Flight Facility
and  scheduled to be launched  mid of 2016 by SpaceX.
The detector includes: four layers of  Si pixels to measure redundantly the particle charge with resolution of 0.2 charge units;
a 0.5 $\lambda_I$-thick carbon target to induce the inelastic interaction of the incoming nuclei; 
 a sampling calorimeter  made of 
20 layers of alternating tungsten plates and scintillating fibers, 
providing energy measurement, particle tracking and trigger; 
top and bottom plastic scintillator counters and a boronated scintillator detector for $e/p$ separation.
While the Si charge detector and the calorimeter
were already used in the balloon campaign and rearranged for ISS-CREAM payload, 
the last two detectors have been newly developed for the space mission,
in order to add sensitivity also to CR electrons.
\subsection{DAMPE (DArk Matter Particle Explorer)}
DAMPE  is a satellite experiment 
promoted by the Chinese Academy of Sciences and built in collaboration with institutions 
from Italy and Switzerland \cite{DAMPE}. The launch is planned on  Dec. 2015. 
The detector consists of four subsystems: plastic scintillator strips which serves 
as  anti-coincidence detector for
photon identification, as well as charge detector for CRs;  
a silicon-tungsten tracker-converter (STK) providing  track reconstruction, photon detection, and CR
charge measurement; 
the BGO imaging calorimeter to measure the incoming particle energy;
 and the neutron detector made of boron-doped plastic plates to detect delayed neutron emerging
from hadron showers in order to improve the $e/p$ discrimination.
DAMPE is designed to measure   electrons and
photons between 5 GeV to 10 TeV with excellent resolution (1.5\% at 100 GeV)
to search for possible DM signatures, and CR nuclei
between 10 GeV and 100 TeV with individual element separation.
The total thickness of the calorimeter is equivalent to 31 $X_0$ and 1.6 $\lambda_I$.
The geometrical acceptance is $\sim$0.3 m$^2$sr
for $e^-$ and $\gamma$, and $\sim$0.2 m$^2$sr for hadrons.
\subsection{CSES (China Seismo-Electromagnetic Satellite)}
CSES is a chinese-italian space mission scheduled to be launched in 2016.
It  aimed at studying  the electromagnetic anomalies from Earth and ionosphere perturbations possibly
associated with earthquakes, in order to develop new technologies to prevent disaster. 
In particular the HEPD (High Energy Particle Detector, developed by the italian group) onboard the satellite will
search for correlations between the precipitation of low energy  electrons
trapped within the Van Allen Belts and earthquakes with Richter magnitude $>$5. 
Besides this scope, HEPD will study the low-energy component of CRs of solar and galactic origin,
 complementing  PAMELA and AMS measurements in the sub-GeV region.
It consists of two layers of plastic scintillators for trigger,  
two planes of double-side Si microstrip detectors to measure the incident particle direction, 
a calorimeter made of plastic and LYSO crystal scintillators, 
and a veto system on the sides.
The total weight is 35 kg and the mechanical size roughly 20$\times$20$\times$40 cm$^3$.
HEPD can measure electrons from 1 to 200 MeV, and protons from 30 to 200 MeV with very good separation,
 an energy resolution <10\% at 5 MeV and an angular resolution of $\sim$8$^\circ$ \cite{CSES}.
\subsection{Proposed experiments}
GAMMA-400  is a next generation gamma-ray space observatory planned for 2023-2025 \cite{GAMMA400}.
Though specifically designed to
measure gamma-ray sources with unprecedented accuracy, the  instrument also 
includes an innovative homogeneous and isotropic calorimeter made of cubic crystals, 
featuring a depth of 54 X$_0$ or
2.5 $\lambda_I$ when detecting laterally incident particles. 
The resulting improved  shower containment and the large acceptance (a few m$^2$sr) would allow to measure
 CR nuclei up to the ``knee'' with very good energy resolution. 

Heavy Nuclei eXplorer (HNX) is a NASA  experiment  
conceived as the successor of SuperTIGER in space \cite{HNX}. 
It exploits complementary active detectors (silicon strips and Cherenkov counters with acrylic and silica-aerogel radiators), 
and passive glass tiles to achieve 50 m$^2$sr geometric factor, required to measure nuclei up to the 
Actinides (Th, U, Pu). 
These are very important CR clocks, 
which can provide information about
the absolute age of ultra-heavy GCRs since nucleosynthesis.
HNX is designed to be accomodated in the DragonLab Capsule for a 2 year mission. 

HELIX (High Energy Light Isotope eXperiment) is a balloon-borne 
experiment  consisting of a  superconducting magnet spectrometer, 
a Ring Imaging Cerenkov Detector and a Time-of-Flight system
optimized to achieve a separation  of 
adjacent light isotopes with  0.25 amu resolution \cite{HELIX}. 
The main goal is to measure with good statistics the $^{10}$Be/$^{9}$Be
abundance ratio 
up to 10 GeV/n, where available data are  lacking. 
This ratio is extremely important to provide information 
on the mean containment lifetime of CRs in the Galaxy and  the diffusion coefficient. 

The General Antiparticle Spectrometer (GAPS) is a large-acceptance balloon-borne experiment 
devoted to indirect DM search by measuring low-energy ($<$1 GeV/n)
CR antideuterons and $\bar{p}$ with a new detection technique, nearly background free \cite{GAPS}. 
Antimatter particles are identified by the 
characteristic X-rays emitted by exotic atoms they form when they are stopped in 
a stack of Si(Li)  detectors, and by the production of a $\pi$ or $p$ star
when subsequently they  annihilate. 
\section{Conclusions}
\label{s:conclusion}
A wealth of new direct CR measurements have been presented at the 34$^{th}$ ICRC.
Among many interesting results, I would like to recall the following: 
\begin{itemize} \itemsep1pt \parskip0pt \parsep0pt
\item[-] AMS precise measurements of the spectral hardening of $p$ and He spectra at $\sim$300 GV, with He harder than $p$;
\item[-] preliminary AMS observation of a hardening also in Li spectrum, not in C one;
\item[-] new AMS B/C data, approaching TeV/n scale with unprecedented precision;
\item[-] Voyager-1 measurements of the unmodulated energy spectra of CR nuclei in the LISM;
\item[-] new SuperTIGER UHGCR data, pointing to OB associations as CR acceleration site; 
\item[-] first detection of a primary CR clock ($^{60}$Fe) from CRIS;
\item[-] $e^\pm$  spectra inconsistent with a single power-law behaviour, with $e^+$  harder than $e^-$;
\item[-] flat behaviour of AMS $\bar{p}/p$ ratio at high energy. 
\end{itemize}
A great theoretical effort is currently underway to explain and interpret these new results, 
which challenge the standard paradigm of CR acceleration and propagation.
A lot of new precise data are foreseen in the years to come
from AMS and several new projects at the horizon.\\\\
{\em I would like to thank the organizers of the conference for inviting me as rapporteur and the local organising committee for the support in collecting and making available talks and papers. I also thank Dr. M. Boezio, Prof. M.H. Israel, Prof. P.S. Marrocchesi, Dr. J.W. Mitchell, and Dr. N. Mori for useful comments during the preparation of the talk and paper. 
} 

\end{document}